\documentclass[useAMS,usenatbib]{mn2e}
\usepackage{amsmath,fleqn,graphicx,amssymb}
\usepackage{multirow}
\renewcommand{\[}{\begin{equation}}
\renewcommand{\]}{\end{equation}}
\def\p{\partial}

\let\boldgrk=\gkvecten
\let\boldgrksc=\gkvecseven

\def\gkthing#1{{\mathchoice%
	{\hbox{{\boldgrk\char#1}}}
	{\hbox{{\boldgrk\char#1}}}
	{\hbox{{\boldgrksc\char#1}}}
	{\hbox{{\boldgrksc\char#1}}}}}

\def\vtheta{\gkthing{18}}

\def\vmu{\gkthing{22}}

\def\vsigma{\gkthing{27}}

{\newif\ifnotend
\notendtrue
\def\veclist{ABCDEFGHIJKLMNOPQRSTUVWXYZabcdefghijklmnopqrstuvwxyz.}
\def\top#1#2.{#1}
\def\tail#1#2.{#2.}
\loop\expandafter\xdef\csname v\expandafter\top\veclist\endcsname%
{{\noexpand\bf\expandafter\top\veclist}}
\edef\veclist{\expandafter\tail\veclist}
\if\veclist.\notendfalse\fi\ifnotend\repeat}
\def\d{{\rm d}}
\def\kB{k_{\rm B}}

\def\df{\textsc{df}}

\def\LCDM{$\Lambda$CDM}
\def\LOSI{\mathrm{LOSI}}
\def\like{\mathcal{L}}
\def\sel{\mathrm{S}}

\def\kpc{\,\mathrm{kpc}}

\def\cut{\mathrm{cut}}

\def\kms{\,\mathrm{km\,s}^{-1}}
\def\masyr{\,\mathrm{mas\,yr}^{-1}}
\def\mas{\,\mathrm{mas}}

\def\msun{\,{\rm M}_\odot}

\def\vlos{{v_\parallel}}
\def\norm{A}
\def\pc{\,\mathrm{pc}}
\def\d{\mathrm{d}}\def\e{\mathrm{e}}
\def\Rc{R_\mathrm{c}}
\def\fracj#1#2{{\textstyle{#1\over#2}}}

\title[Determining the Milky Way potential]
{Analysing surveys of our Galaxy -- II. Determining the potential}

\author[P.~J.~McMillan and J.~J.~Binney]{
  Paul~J.~McMillan\thanks{E-mail: p.mcmillan1@physics.ox.ac.uk} and James~J.~Binney
  \\  
  Rudolf Peierls Centre for Theoretical Physics, 1 Keble Road,
  Oxford, OX1 3NP, UK
}

\begin{document}
\maketitle

\begin{abstract}
 We consider the problem of determining the Galaxy's gravitational potential
from a star catalogue.  We show that orbit-based approaches to this problem
suffer from unacceptable numerical noise deriving from the use of only a
finite number of orbits.  An alternative approach, which requires an ability
to determine the model's phase-space density at predetermined positions and
velocities, has a level of numerical noise that lies well below the intrinsic
uncertainty associated with the finite size of the catalogue analysed.  A
catalogue of $10\,000$ stars brighter than $V=17$ and distributed over the
sky at $b>30$ degrees enables us to determine the scaleheight of the disc
that contributes to the potential with an uncertainty below $20\pc$ if
the catalogue gives proper motions, line-of-sight velocities and parallaxes
with errors typical of the Gaia Catalogue, rising to $36\pc$ if only proper
motions are available. The uncertainty in the disc's scalelength is
significantly smaller than $0.25\kpc$.
\end{abstract}


\section{Introduction} \label{sec:intro}

Very large resources are currently being devoted to surveys of our
Galaxy, both from the ground \citep[e.g. APOGEE, RAVE and
Gaia-ESO:][]{SDSS-III,RAVE1_short,GaiaESO} and from space
\citep[e.g. Gaia:][]{GAIA01}. These surveys are being undertaken in the
expectation that they will reveal the current kinematic and chemical
structure of our Galaxy and how our Galaxy was assembled. The latter
is a pivotal question for cosmology since our Galaxy is typical of the
galaxies that dominate star formation in the contemporary Universe,
and the consensus \LCDM\ cosmology has given us a moderately clear
view of how such galaxies formed.

Since stars and dark matter orbit freely in the Galaxy's gravitational
potential, our understanding of the Galaxy can never be better than
our knowledge of its gravitational potential $\Phi(\vx)$. Since even
at the Sun we are unable to measure the density of dark matter, $\Phi$
cannot be determined from Poisson's equation, but must be constrained
by measuring the motions of objects that we can see -- in practice
stars and interstellar gas.  Historically, the crucial data have been
the circular speed in the plane, $v_{\rm c}(R)$, estimated from the
line-of-sight velocities $\vlos$ of interstellar gas
\citep{Malhotra94}, the local density $\rho(R_0)$ estimated from the
kinematics of nearby stars \citep{Creze98,HolmbergFlynn}, the local
surface density $\Sigma_{1.1}$ estimated from observations of stars a
few hundred parsecs away from the plane \citep{KuGi89}, and the proper
motion of Sagittarius A* at the Galactic centre \citep{ReBr04,Giea09}.
One can find models of the Galactic potential which fit all of these
constraints \citep*[e.g.][]{WDJJB98:Mass,KlZhSo02,PJM11:mass}, but
unfortunately they do not constrain the density of dark matter very
strongly.  The data that are now available to us, or will shortly
become available, open new horizons in the determination of $\Phi$.

The questions we address are:
\begin{itemize}
\item given a star catalogue, what is the best way to constrain
$\Phi$?
\item When we proceed in the optimal way, how strongly do data of the
  type that will shortly be available constrain $\Phi$?
\end{itemize}

The data available from surveys of the Galaxy fundamentally differs
from that generally available for external galaxies. It is practical
to determine not just the position on the sky and line-of-sight
velocity of a star, but also proper motions and distance from the Sun
with a useful degree of accuracy. However, there is a strong bias in
which stars are observed by the survey -- those sufficiently close to
the Sun's position in the Galaxy and which lie within the selection function
of the survey. For observations of external galaxies there is no such
bias created by the Sun's position, and full surface brightness
profiles can be found, but the velocity information is generally
limited to binned line-of-sight velocity distributions. Ensuring that
models can accurately represent data of the precision available for
stars in the Galaxy is a significant problem.

A compelling argument can be made that the availability of models of
sufficient sophistication is the key to extracting science from a star
catalogue \citep[][henceforth Paper I]{PJMJJB12}. One must
discriminate between various model Galaxies by inferring their a
posteriori probabilities, given the catalogue. The simplest possible
dynamical model Galaxy is defined by the pair of functions $(f,\Phi)$,
where $f(\vx,\vv)$ is the distribution function (\df).

When estimating $\Phi$ an unavoidable assumption is that the Galaxy is in a
steady state: without this assumption the observed kinematics of any tracer
objects are consistent with any potential; it is only by assuming that $\Phi$
is deep enough to prevent a tracer population expanding, and not so deep as
to cause contraction in the next dynamical time that we can constrain $\Phi$.
The assumption that our tracers are in dynamical equilibrium permits us to
invoke the strong Jeans theorem and conclude that the
\df\ $f$ of our tracer population is a function $f(I_1,I_2,\ldots)$ of
whatever isolating integrals $\Phi$ admits. In the 1970s numerical
experiments revealed that typical galaxy potentials admit three independent
isolating integrals, for example, energy $E=\fracj12v^2+\Phi$, the component
of angular momentum $L_z$ about the potential's symmetry axis and a ``third
integral'' $I_3$, which controls the division of energy in excess of the
energy $E_c(L_z)$ of a circular orbit of angular momentum $L_z$ between
oscillation in radius and oscillation perpendicular to the potential's
equatorial plane. 

Any function of isolating integrals is itself an isolating integral, so there
is considerable flexibility in the choice of arguments of the \df. We have
argued elsewhere \citep{PJMJJB08,JJB10} that there are compelling
reasons to choose the actions $J_r$, $J_\phi\equiv L_z$ and $J_z$ as our
isolating integrals. So we work with these integrals here and take a
model galaxy to be the pair of functions $(f(\vJ),\Phi(\vx))$. However,
working with alternative integrals would not change our conclusions in any
essential way; alternative integrals would simply make the computations
harder and less transparent. In particular, our arguments apply to models of
the  type that are most widely used in studies of external galaxies:
\cite{Sc79} models. These models comprise a potential $\Phi$ and an
orbit library, each element of which is the time series of phase-space
coordinates $(\vx(t),\vv(t))$ obtained by integrating the equations of motion
in $\Phi(\vx)$ for a particular initial condition, and a weight $w\ge0$ with
which that time series is employed in the model. In effect these models use
the initial conditions of integrations as isolating integrals, and the weights
$w$ are surrogates for the value of the \df\ on the given initial conditions.

In general terms the procedure for modelling a catalogue is to
determine the probability of the data given some pair $(f,\Phi)$ and
then to use Bayes' theorem to convert the probability of the data into
the probability of the pair. Ideally, the probabilities of every
plausible pair $(f,\Phi)$ would be determined, but in practice one has
to be content with a search for a limited number of more likely
pairs. The problem is made computationally tractable by considering
each candidate potential in turn, and then finding the most probable
companion \df. In Paper I we showed that when a catalogue of $10\,000$
stars is constructed in a known potential from a \df\ of a given
functional form, the \df\ can be recovered to good precision from the
catalogue. In this paper we explore our ability to determine the
correct potential by repeating the \df-fitting step for a series of
potentials and identifying the potential that yields the largest
likelihood for the catalogue.

We show that this problem cannot be efficiently solved by any orbit-based
technique such as N-body modelling, Schwarzschild modelling, or torus
modelling of the type used in Paper I. We show further that the problem can
be solved if we have available expressions for the isolating integrals as
functions of the conventional phase-space variables, for example
$\vJ(\vx,\vv)$. This second approach was adopted by
\cite{YSTea13}, and we extend it to include both observational errors
and realistic selection effects, and to exploit the more powerful approach to
the determination of actions of \cite{JJB12:Stackel}.

The outline of the paper is as follows. In Section~\ref{sec:aa} we
describe the catalogues that we consider, the models that we compare
them to, and the tools used to do the comparison. In
Section~\ref{sec:approaches} we explain the two competing methods we
use to analyse the data. In Section~\ref{sec:problem} we show that
orbit-based methods are ill-suited to this analysis. In
Section~\ref{sec:solution} we demonstrate methods using expressions
for $\vJ(\vx,\vv)$ that are capable of successfully determining the true
potential.

\section{Theoretical framework}
\label{sec:aa}

Three actions $J_i$ and three conjugate angle coordinates $\theta_i$
provide exceptionally convenient coordinates for objects orbiting in a
stationary gravitational potential $\Phi$.  The actions are conserved
quantities and the angles increase linearly with time, $\theta_i(t) =
\theta_i(0)+\Omega_i(\vJ)t$, where $\Omega_i$ is a frequency.

Thus $\vJ$ labels an orbit and
$\vtheta$ specifies a point on that orbit. The usual phase space
coordinates $\vx,\vv$ are $2\pi$-periodic in each angle coordinate
$\theta_i$. Angle-action coordinates have the convenient property that 
\[
\left|\frac{\p (\vtheta,\vJ)}{\p (\vx,\vv)}\right| = 1,
\]
 so it is simple to relate a density in angle-action
space to the phase-space density $f(\vx,\vv)$.

The relationship $(\vtheta,\vJ) \leftrightarrow (\vx,\vv)$ depends
upon the gravitational potential $\Phi$. Unfortunately analytical
expressions for this relationship are only known for a very limited
set of potentials. In recent years a great deal of effort has gone
into developing numerical approximations to this
relationship \citep[][]{PJMJJB08,JJB10,JJBPJM11,Sa12,JJB12:Stackel}. In this
paper we first use the machinery described in \cite{PJMJJB08} that yields
$(\vx(\vtheta,\vJ),\vv(\vtheta,\vJ))$ and then the machinery described in
\cite{JJB12:Stackel}, which gives the inverse transformation
$(\vtheta(\vx,\vv),\vJ(\vx,\vv))$.

\subsection{Torus modelling} 

Torus modelling \citep[][and references therein]{PJMJJB08} is a method which,
for a single value $\vJ$ in a given potential $\Phi$, provides (through a
numerical minimisation) an expression for the phase-space coordinates
$(\vx,\vv)$ in terms of $\vtheta$. Thus it tells us the complete phase-space
structure of the specified orbit $\vJ$. We refer to this model of an orbit as
a ``torus'' because the three-dimensional surface mapped out in phase space
as the $\theta_i$ vary over their full range $(0,2\pi)$ is isomorphic to a
3-torus.

Thus with this method it is easy to find $(\vx,\vv)$ in terms of
$(\vtheta,\vJ)$, but
far harder to find $(\vtheta,\vJ)$ given $(\vx,\vv)$ \citep[though
it can be done iteratively, e.g.][]{PJMJJB08}.

Torus modelling is best understood as an extension of Schwarzschild modelling
\citep{Sc79} in which time series are replaced by orbital tori. This
replacement brings a number of advantages \citep{JJBPJM11}. Torus modelling
was the method used to in Paper I, and has also been used in modelling the
Hyades moving group \citep{PJM13:Hyades} and to show how one can disentangle
the history of a disrupted satellite object that we observe as debris in the
Solar neighbourhood \citep{PJMJJB08}.

\subsection{St\"ackel approximation}

\cite{JJB12:Stackel} introduced an algorithm for calculating the actions of
stars with known phase-space coordinates in axisymmetric potentials. It is
based upon the approximation that in the region probed by a given orbit, the
potential of interest does not differ greatly from a St\"ackel potential
\citep[e.g.][\S3.5.3]{deZ85,GDII}. With this assumption it is possible to estimate the
radial and vertical actions $J_r$ and $J_z$ at any point $(\vx,\vv)$ from
one-dimensional integrals over coordinates that form a system of confocal
ellipsoidal coordinates. Thus when this apparatus is used, it is easy to
obtain $(\vtheta,\vJ)$ from $(\vx,\vv)$ but hard to proceed in the opposite
direction. 

This approximation gives values of $\vJ$ for typical orbits in the
thin disc which are around a factor of $4$ more accurate than those
found by the ``adiabatic approximation'' which has been used for the
same purpose \citep[e.g.][]{JJB10,YSTea13}. It provides an even
greater improvement in accuracy for the orbits of many thick-disc
stars, to which the adiabatic approximation does not apply. It has
been used to fit \df s to observational data for the Solar
neighbourhood, given assumed gravitational potentials
\citep{JJB12:dfs}.

\subsection{Distribution functions}

The distribution functions used in this paper are all based upon the
``quasi-isothermal'' \df\ \citep{JJBPJM11}. Here we modify the notation used
previously in two respects: (i) we change the normalisation of $f$ so when
integrated over all phase space it produces unity, and (ii) we replace the
parameter $q$ by $R_\sigma\equiv R_\d/q$, where $R_\sigma$ is the radial
scale on which the velocity dispersions decline with increasing radius, and
$R_\d$ is the conventional scalelength of the approximately isothermal disc.
 \begin{equation}\label{totalDF}
f(\vJ)\equiv{\Omega\nu\Sigma\over2\pi^2 M\sigma_r^2\sigma_z^2\kappa}
\bigg|_{\Rc}
\cut(L_z)\;{\rm e}^{-{\kappa J_r/\sigma_r^2}}\,{\rm e}^{-{\nu J_z/\sigma_z^2}},
\end{equation}
with $\vJ\equiv(J_r,J_z,L_z)$. 
Here $\Omega(L_z)$ is the circular frequency for angular momentum
$L_z$, $\kappa(L_z)$ is the radial epicycle frequency and $\nu(L_z)$
is its vertical counterpart.  $\Sigma(L_z)=\Sigma_{0}\e^{-\Rc/R_\d}$
is the (approximate) radial surface-density profile, where $\Rc(L_z)$
is the radius of the circular orbit with angular momentum $L_z$, and
$M = 2\pi\Sigma_0 R^2_\d$ is a constant included to ensure that 
\[
\int\d^3\vJ\,f(\vJ)=1.
\]
The factor $\cut(L_z)$ is included to ensure that we do not have
equal numbers of stars rotating in each direction. We use 
\[
\cut(L_z) = {\textstyle{1\over2}}\left[1+\tanh(L_z/L_0)\right]
\]
where the value of
$L_0$ is unimportant in this study provided it is small compared to
the angular momentum of the Sun. We hold it fixed at
$L_0=10\kpc\kms$.
The
functions $\sigma_z(L_z)$ and $\sigma_r(L_z)$ control the vertical and
radial velocity dispersions. We adopt
 \begin{eqnarray}\label{eq:sigmas}
\sigma_r(L_z)&=&\sigma_{r0}\,{\rm e}^{(R_0-\Rc)/R_\sigma}\nonumber\\
\sigma_z(L_z)&=&\sigma_{z0}\,{\rm e}^{(R_0-\Rc)/R_\sigma},
\end{eqnarray}
where $\sigma_{r0}$ and $\sigma_{z0}$ are parameters that are set to
values close to the radial and vertical velocity dispersions at the
Sun. The observed insensitivity to radius of the scaleheights of
extragalactic discs suggests $R_\sigma\sim 2R_\d$, if $R_\d$ is the
scalelength of the disc that dominates the potential. To simplify
calculations we hold $R_\sigma=R_\d/0.45$ (as in Paper I).

\cite{JJB12:dfs} showed that by superposing a large number of
quasi-isothermal \df s, one can obtain a model that is consistent with
the local stellar density and velocity distribution as revealed by the
Geneva-Copenhagen survey \citep{GCS09}. In this study we,
as in Paper I, restrict ourselves to simple two-disc models in
order to provide some straightforward demonstrations of the principles
involved. These are of the form
\[
f(\vJ) = (1-\lambda)\, f_{\rm thin}(\vJ) + \lambda\, f_{\rm thick}(\vJ)
\]
 with $f_{\rm thin}$ and $f_{\rm thick}$ of the form given in
equation~\ref{totalDF}, and $\lambda$ the fraction in the thick disc. Extending
this work to more complicated \df s is, in principle, straightforward.

\subsection{Numerical details of models used in this study}\label{sec:models}

The models we use to test our analysis techniques are discrete
realisations obtained by sampling a \df\ that has two quasi-isothermal
discs with parameters as listed in Table~\ref{tab:df}. We refer to
this \df\ as $f_{\rm true}$. This is identical to one of the \df s
used in Paper I and it is sampled using the torus machinery as
described in Paper I.

In all cases the model is constructed in the ``convenient'' Milky Way
potential given by \cite{PJM11:mass}, which we will refer to as the
``Mc11'' potential. This is an axisymmetric model,
in which the potential is assumed to be produced by a Galactic bulge,
thin and thick exponential discs, and a spherical \cite*{NFW96}
halo. The density of the bulge is
\begin{equation} \label{eq:bulge}
  \rho_\mathrm{b}=\frac{\rho_{\mathrm{b},0}}{(1+r^\prime/r_0)^\alpha}\;
  \textrm{exp}\left[-\left(r^\prime/r_{\mathrm{cut}}\right)^2\right],
\end{equation}
where, in cylindrical coordinates,
\begin{equation}
  r^\prime = \sqrt{R^2 + (z/q)^2}
\end{equation}
with $\alpha=1.8$, $r_0=0.075\mathrm{kpc}$, $r_{\mathrm{cut}}=2.1\mathrm{kpc}$,
and axis ratio $q=0.5$; the densities of the two discs are of the form 
\begin{equation} \label{eq:disc} 
  \rho_\mathrm{d}(R,z) =
  \frac{\Sigma_{\mathrm{d},0}}{2z_\mathrm{d}}\;\exp\left(-\frac{|z|}
{z_\mathrm{d}}-\frac{R}{R_\mathrm{d}}\right),
\end{equation}
with scaleheight $z_\mathrm{d}$, scalelength $R_\mathrm{d}$ and central surface
density $\Sigma_{\mathrm{d},0}$; and the density of the halo is of the form 
\begin{equation} \label{eq:NFW} \rho_{\mathrm{h}} =
  \frac{\rho_{\mathrm{h},0}}{x\,(1+x)^2},
\end{equation}
where $x=r/r_{\mathrm{h}}$, with $r_\mathrm{h}$ the scale-radius. The
parameters of the Mc11 model are shown in
Table~\ref{tab:conv}. 
\begin{table*}
  \begin{tabular}{ccccccccc} \hline
    $\Sigma_{\mathrm{d},0,\mathrm{thin}}$ & $R_{\mathrm{d},\mathrm{thin}}$
    & $z_{\d,{\rm thin}}$ & $\Sigma_{\mathrm{d},0,\mathrm{thick}}$ &
    $R_{\mathrm{d},\mathrm{thick}} $ & $z_{\d,{\rm thick}}$ & 
    $\rho_{\mathrm{b},0}$ & $\rho_{\mathrm{h},0}$ & $r_\mathrm{h} $\\ \hline
    $753.0\msun\pc^{-2}$ & $3\kpc$ & $0.3\kpc$ & $182.0\msun\pc^{-2}$
    & $3.5\kpc$ & $0.9\kpc$ &
    $94.1\msun\pc^{-3}$ & $0.0125\msun\pc^{-3}$ & $17\kpc$  \\ \hline
  \end{tabular}
  \caption{
Parameters of the Mc11 model used in the construction of the particle
model. Note that the potentials with differing scaleheights used in
this study have identical parameters except the two disc scaleheights
(which are held with $z_{\d,{\rm thick}}/z_{\d,{\rm thin}}=3$, with
the quoted scaleheight always being that of the thin disc). The
potentials with varying $R_d$ have different parameters, listed in
Table~\ref{tab:Rd}. 
}
\label{tab:conv}
\end{table*}

We also consider a range of other potentials, to show how well our
analysis techniques work when asked to compare the likelihoods of
competing plausible potentials.  We perform a number of tests in which
all parameters are held constant at those of Mc11, except the disc
scaleheights -- in these cases the ratio of the two disc scaleheights
is held constant, and the quoted value is that of the thin-disc
scaleheight. 

We also perform tests in which the disc scalelengths are varied. Again we
hold the ratio of the two disc scalelengths constant. However, now leaving
the other parameters of the potential unchanged would lead to substantial
changes in the properties of the model (for example in the circular speed),
so we constrain the parameters of each potential in the same way as in
\cite{PJM11:mass}, with the Sun's position and the disc scaleheights being
held constant.

The constraints on the potentials employed include the proper motion of Sgr
A* \citep{ReBr04}. Given the Sun's Galactocentric distance ($8.5\kpc$) and
peculiar motion with respect to the local standard of rest, \citep*{SBD10}
the proper motion of Sgr A* strongly constrains the local circular speed. We
also fit the potentials to the terminal velocity of the ISM at $30^\circ < l
< 90^\circ$ \citep{Ma94,Ma95} and to observed maser sources
\citep[e.g.][]{PJMJJB10:masers} which constrain the shape of the circular-speed
curve, and to the vertical force $1.1\kpc$ from the plane at the Solar radius
\citep{KuGi91}. For full details of the constraints applied, see
\cite{PJM11:mass}.

Constructing each potential in this way ensures that we are comparing
potentials that are all good fits to existing basic kinematic data,
which are currently state-of-the-art constraints on the Galactic
potential. The results of this analysis therefore show how these
dynamical models can increase our knowledge of the potential.

The parameters of these potentials are given in the appendix,
Table~\ref{tab:Rd}, and we refer to them in the text by their thin disc
scalelengths.

\begin{table}
  \caption{Parameters of the \df\ used to construct our catalogues,
    $f_{\rm true}$. The fraction in the thick disc component is $\lambda=0.23$.
  }\label{tab:df}
  \begin{center}
    \begin{tabular}{l|ccc} \hline
      Disc & $R_\d$ & $\sigma_{r0}$ & $\sigma_{z0}$ \\
      & $(\hbox{kpc})$ & $(\kms)$  & $(\kms)$\\
      \hline
      Thin & 3.0 & 27 & 20 \\
      Thick& 3.5 & 48 & 44 \\ \hline
    \end{tabular}
  \end{center}
\end{table}

We consider catalogues of ``observations'' of the discrete realisations. In
general, a catalogue of $N$ stars gives accurate values of the Galactic
coordinates $(b,l)$, values of apparent magnitude $m$, colour $V-I$, and
line-of-sight velocity $\vlos$ that have moderate errors, and values of the
parallax $\varpi$, proper motion $\vmu$, surface gravity $g$ and metallicity
$Z$ that are probably significantly in error. We group the variables into two
sets, the basic variables
\[\label{eq:defsu}
\vu\equiv(b,l,m,\varpi,\vmu,\vlos)
\]
and additional astrophysical variables
\[\label{eq:s}
\vs\equiv (V-I,g,Z).
\]
Note that $\vu$ has seven components, effectively a star's phase-space
coordinates $(\vx,\vv)$ and its apparent magnitude $m$. For now we
neglect interstellar extinction. Then a star's absolute magnitude
$M$ is effectively specified by $\vu$ because its distance is
fixed by $\vx$.

\begin{figure}
  \centerline{\resizebox{.9\hsize}{!}{\includegraphics[angle=270]{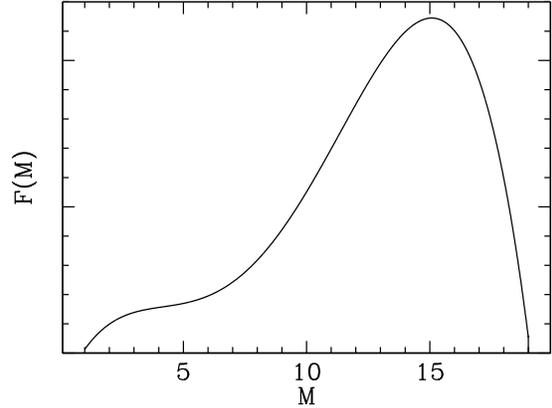}}}
  \caption{Luminosity function given by equation~(\ref{eq:LF}), and
    used in all tests.}
  \label{fig:LF}
\end{figure}

As in Paper I, we restrict ourselves to the case of a
single stellar population. This assumption ensures that there are no
correlations between stellar type and kinematics: the distribution of
stars in phase space is independent of their luminosities, colours,
metallicities, etc. In this case we can confine discussion to the
components of $\vu$ and neglect $\vs$.  We further assume that the
luminosity function $F(M)$ is known to be
\begin{equation} \label{eq:LF} F(M) \propto \left\{
    \begin{array}{ll} 
      -14.9 + 21\,M -5.4\,M^2 \\
      \qquad+ 0.59\,M^3 -0.019\,M^4 & \hbox{ for } 1<M<19 \\
      0 & \mathrm{otherwise}, \\ 
    \end{array}
  \right.
\end{equation}
 which is a simple polynomial approximation to the general $V$-band
luminosity function described in \cite{GA}, Table 3.16.  This function is
plotted in Figure~\ref{fig:LF} and satisfies the normalisation condition
 \[
1=\int_{-\infty}^\infty\d M\,F(M) .
\]
Throughout this paper we use the notation $F(m,s)$ to mean the
luminosity function of an apparent magnitude $m$ at a heliocentric
distance $s$ where, as we neglect extinction,\footnote{Here and
  throughout this paper we use $\log$ to mean $\log_{10}$.}
\[
F(m,s) \equiv F(m-5\log(s/10\pc)) .
\]

We consider three catalogues of ``observations'' that give us information on
the variables $\vu^\alpha$ for each star $\alpha$. In each case the catalogue
contains $N_\alpha = 10\,000$ stars, all at Galactocentric latitude
$b>30^\circ$ and with apparent magnitude $m<17$. We assume that no other
selection effects bias the catalogue. For each observed star we assume that
the quoted Galactic coordinates $(b,l)$ are exact, as is the apparent
magnitude $m$ (this is equivalent to the statement that the uncertainty in
$m$ is much smaller than the scale on which $F(M)$ varies). We then have: 
\begin{itemize}
\item A catalogue with measurements of parallax with uncertainty
  $\sigma_\varpi=0.2\mas$, measurements of line-of-sight velocity with
  uncertainty $\sigma_\parallel=5\kms$, and measurements of proper
  motion with uncertainty (in each direction) of $\sigma_\mu =
  0.2\masyr$.
\item A catalogue with parallax and proper motion measurements with
  the same uncertainty as previously, but with no measurement of the
  line-of-sight velocity.
\item A catalogue with proper motion measurements with the same
  uncertainty as previously, but with no line-of-sight velocity or
  parallax measurements. 
\end{itemize}

To convert values from Galactocentric coordinates to Heliocentric
coordinates $\vu$ (to create this catalogue), or vice versa (to
analyse the catalogue) we need to assume a position and velocity for
the Sun. In all cases we assume that the Sun is at a Galactocentric
radius $R_0=8.5\kpc$ moving with the peculiar velocity found by 
\cite*{SBD10} with respect to the local standard of rest in the currently
hypothesised potential. 

\begin{figure}
\centerline{
\resizebox{1.1\hsize}{50mm}{\includegraphics[0mm,140mm][220mm,250mm]{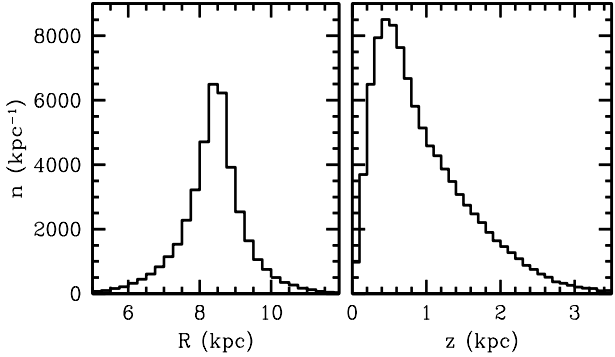}}
}
\caption{Histograms of the number of stars in the catalogue as a
  function of Galactocentric radius (left) and distance from the
  Galactic plane (right).} \label{fig:distrib}
\end{figure}

In Figure~\ref{fig:distrib} we show histograms of the number of stars
in the catalogue as a function of Galactocentric $R$ and $z$. We use
the true positions of the stars to produce these histograms, rather
than adding any uncertainties. 

\section{Assessing a model likelihood: two approaches}
\label{sec:approaches}

\subsection{General notation}

Before describing our new method of solving the problem of determining
model likelihoods, we explain our notation (which is
very similar to that used in Paper I).
We assume that the errors in the observed quantities are independent
and can be modelled by Gaussian probability distributions
 \[
G(u,\overline{u},\sigma)\equiv{1\over\sqrt{2\pi\sigma^2}}\e^{-(u-\overline{u})^2/2\sigma^2}.
\]
 We attach primes to the true values of measured quantities to
distinguish them from the measured values. For brevity we use the
notation
 \[
G_i^j(\vu^\alpha,\vu',\vsigma^\alpha)\equiv\prod_{k=i}^jG(u^\alpha_k,u_k',\sigma^\alpha_k).
\]

We  find the likelihood $\mathcal{L}$ of a model as a
product over stars of the probabilities of measuring the values $\vu^\alpha$
given the model:
 \begin{eqnarray}\label{eq:uprime}
  \mathcal{L} &=& \prod_\alpha \mathcal{L}^\alpha_\ast \equiv \prod_\alpha 
  P(\vu^\alpha|{\rm Model}) \nonumber\\
  & = & \prod_\alpha \int\d^7\vu'\,  G_1^7(\vu^\alpha,\vu',\vsigma^\alpha) \times
  P(\vu'|{\rm Model}).
\end{eqnarray}
Any quantity that is not given in the catalogue (such as $\vlos$ in
two of our catalogues) can be considered to have a sufficiently large
$\sigma$ that the Gaussian density is effectively constant for all
relevant values of the variable.

The probability $P(\vu'|{\rm Model})\,\d^7\vu'$ is the
probability that \emph{a randomly chosen star in the catalogue} has the true
values $\vu'$, for a particular model. Specifically
$P(\vu')$ is given by
 \[\label{eq:pumod}
P(\vu'|{\rm Model}) = \norm\, \sel(\vu') F(M) f(\vx,\vv)
\left| {\partial(M,\vx,\vv)\over\partial(\vu')} \right| ,
\]
 where the selection function $\sel(\vu')$ is the probability that if a
star with observables $\vu'$ exists it will be included in the
catalogue, and the normalisation factor $\norm$  depends
on the \df, the luminosity function and the survey selection effects
via the equation
\[\label{eq:A}
1/\norm = \int\d^3\vx\,\d^3\vv\,f(\vx,\vv)\int  \d M\,\sel(\vx,\vv,M) F(M) .
\]
As we assume that the \df\ $f(\vx,\vv)$ is that of an equilibrium
dynamical model, we have
\[
\int\d^3\vx\,\d^3\vv\, f(\vx,\vv)  = (2\pi)^3\int \d^3\vJ\, f(\vJ) = 1,
\]
 where the integrals are over all phase space and action space, respectively.

Our catalogue contains stars at Galactic latitudes $b>b_{\rm lim}$  and apparent
magnitude $m<m_{\rm lim}$ with $b_{\rm lim}=30^\circ$ and 
$m_{\rm lim}=17$. We assume that, other than these limits, the
probability of a star entering the catalogue is independent of its
properties, i.e.
\[\label{eq:sel}
\sel(\vu') =\left\{ 
  \begin{array}{ll} {\rm const} & \hbox{ for } b>b_{\rm lim}, m<m_{\rm lim} \\
    0 & \hbox { otherwise. } \\
    \end{array} \right. 
\]
Note that as this constant then appears in both the numerator and
denominator (implicitly) of eq.~(\ref{eq:pumod}), we can ignore it.

Since the potential $\Phi$ determines the relationship between $f(\vx,\vv)$
and $f(\vJ)$, the normalisation factor $\norm$ (eq.~\ref{eq:A}) is a
function of $\Phi$ in addition to $f(\vJ)$, so we might write
$P(\vu'|{\rm Model})\equiv P(\vu'|f,\Phi,F,\sel)$. 

Formally, the best way to constrain the potential is to marginalise
over all possible \df s \citep{Ma06,monkeys2}. While our use of \df s
that are functions of $\vJ$ opens up this exciting possibility, we do
not attempt it in this study. \cite{YSTea13} marginalised over the
parameters of a single pseudo-isothermal \df. This relatively
low-dimensional marginalisation still required at least 1 to 2 orders
of magnitude longer than simply finding the maximum likelihood for a
\df\ of the assumed form in a given potential. They found that this
marginalisation made little difference to their results (Ting,
priv. comm.). In this study we only consider the maximum likelihood
for a \df\ of the assumed form in the chosen $\Phi$. This is very much
like the approach used by Schwarzschild modellers.

\subsection{Evaluating with $\vx(\vtheta,\vJ)$}

We first consider the approach to likelihood evaluation that was successfully
employed in Paper I. This is founded on a library of tori, each of which
yields $(\vx(\vtheta,\vJ),\vv(\vtheta,\vJ))$. The problems we encounter would
be at least as serious if we were to replace the torus library by an orbit
library of the type introduced by \cite{Sc79}.  Our description of the use
of tori will be brief; a reader looking for more detail should read Sections
4~\&~5 of Paper I. 

A torus provides complete knowledge of the orbit with actions $\vJ$ in a
given potential . We therefore convert the integrals in
equations~(\ref{eq:uprime})~\&~(\ref{eq:A}) into integrals over
$(\vtheta,\vJ)$. The normalisation factor $\norm$ (eq~\ref{eq:A}) is rewritten
as
 \[\label{eq:ATorus}
1/\norm = \int f(\vJ) \phi(\vJ)\,\d\vJ ,
\]
where
\[\label{eq:phi}
\phi(\vJ)\equiv\int\d^3\vtheta\,\int\d m\,F(m,s)
\,\sel(\vx,\vv,m).
\]
 Note that $\phi(\vJ)$ depends on the potential. Regions of action space with
$\phi(\vJ)=0$ (in a given potential) do not contribute to the calculation and
can be ignored -- though the analytic forms we use for the \df\ do make
predictions of the \df\ outside the survey volume, which can then be tested
by further data.

As in Paper I we use the approximations that the measurements of $b$, $l$
and $m$ are exact to perform three of the seven  integrals in
equation~(\ref{eq:uprime}), leaving integrals over $\vJ$ and $s'$ (i.e. along the
line of sight):
\begin{eqnarray} 
  \mathcal{L}_\ast^\alpha & = & \norm\! \int\!\d^3\vJ  f(\vJ)\int\!\d^3\vtheta \,\d
  M\,
  G_1^7(\vu^\alpha,\vu',\vsigma^\alpha) F(M) \sel(\vu') \nonumber\\
  & = & \norm \int\d^3\vJ  \int \d s' \left|
    {\p(\vtheta)\over\p(b,l,s')}\right|
  \nonumber\\
  & &  \qquad\qquad \qquad\;\; \times\; 
G_4^7(\vu^\alpha,\vu',\vsigma^\alpha)F(m,s') \sel(\vu'),
\end{eqnarray}
 where $\vu'(\vtheta,\vJ,m)$ and $\vtheta$ is  now a function of $s'$ because
$(b,l)$ are known.
 
 The principle of Monte-Carlo integration is now invoked to convert these 
two integrals into sums over points
$\vJ_k$ that have been selected  with a sampling density $f_S(\vJ)$ that will
be described below. Then we have
\[ \label{eq:calcTorus} 
\mathcal{L}_\ast^\alpha  = \sum_k
  \frac{f(\vJ_k)}{f_S(\vJ_k)}
  \mathrm{LOSI}_{k,\alpha}\bigg/ \sum_k
  \frac{f(\vJ_k)}{f_S(\vJ_k)}\phi_k,
\]
 where we have used the notation $\phi_k\equiv\phi(\vJ_k)$, and introduced
the line-of-sight integral for a given $\vJ_k$ and observation $\alpha$
\[\label{eq:LOSI}
\mathrm{LOSI}_{k,\alpha} = \int \d s' \left|
  {\p(\vtheta)\over\p(b,l,s')}\right|_k F(m,s')
G_4^7(\vu^\alpha,\vu',\vsigma^\alpha) \sel(\vu').
\]
 The computation of these $N_k\times N_\alpha$ line-of-sight integrals
dominates the computing budget for these calculations. Note that
given our choice of selection function (eq.~\ref{eq:sel}), we clearly
have $\sel(\vu')={\rm const}\neq 0$ in all cases. The Jacobian
$\left|{\p(\vtheta)\over\p(b,l,s')}\right|_k$ can be found using the
torus machinery \citep{JJBPJM11}, and is closely related to the
density of the orbit.

This approach allows us to reuse the values $\mathrm{LOSI}_{k,\alpha}$ and
$\phi_k$ that we have determined for the calculation of $\like$ for many \df
s in a given potential, but note that both depend critically on the
relationship between $\vu'$ and $(\vtheta,\vJ)$, which depends on the
potential, so they cannot be reused when we move to a new potential.

\subsection{Evaluating with $\vJ(\vx,\vv)$}\label{sec:EJxv}

Given that we have expressions for $\vJ$ in terms of $(\vx,\vv)$, which we
find using the St\"ackel approximation, we can evaluate the normalisation
factor $\norm$ directly from equation (\ref{eq:A}) rather than its
angle-action reformulation (\ref{eq:ATorus}).  We use the sampling density
$f_S(\vx,\vv)$ described below to ensure the points are concentrated
where the integrand is largest.  Then we have to evaluate the
Monte-Carlo sum
 \[\label{eq:AStack}
1/\norm = \frac{1}{N}\sum_{k=1}^N {f(\vJ(\vx_k,\vv_k))\over f_S(\vx_k,\vv_k)}
\int\d  m\,F(m,r_k) \sel(\vx_k,\vv_k,m).
\]
 Since $1/\norm$ is a factor of every star's likelihood $\like^\alpha_\ast$, the
total likelihood $\like\propto \norm^{-N_\ast}$. Consequently, if we have a
fractional error of $\delta$ in the value of $\norm$, the error in
$\log\like$ is $\sim0.43\,N_\ast\,\delta$, so for our catalogues of $10\,000$
stars, we would require an uncertainty of $\sim0.02$ per cent in $\norm$ to
limit the uncertainty in $\log\like$ to order unity.  Numerical experiments
to be described below reveal that with $4\times10^6$ sample points the
uncertainty in $\norm$ is $\sim0.4$ per cent, $\sim 10^9$
sample points are required to determine the $\log\like$ to order unity. This
is a very challenging requirement!

Fortunately the absolute value of $\like$ is not important. What
matters is the ratio of two given values of $\like$. We can minimise
the numerical noise in this ratio by fixing the points $(\vx_k,\vv_k)$
at which we evaluate the Monte-Carlo sum (eq~\ref{eq:AStack}) for
all models, independently of the potential.

The great advantage of using $\vJ(\vx,\vv)$ rather than $\vx(\vtheta,\vJ)$ is
that it is now possible to derive a Monte-Carlo sum for the likelihood in
eq.~(\ref{eq:uprime}) that runs over observable points $\vu'$ that are
determined by the observations and, again, are independent of the potential.  
Moreover we will be able to arrange that each star's sample points will be
concentrated within its error ellipsoid, rather than distributed
regardless of where the star is observed.

We start by writing the integral associated with the $\alpha$th star as
\begin{eqnarray}\label{eq:LalphaJx}
\like_\ast^\alpha & = & \norm\int\d^7\vu'
\left|\frac{\p(\vtheta,\vJ,M)}{\p(\vu')}\right| \nonumber\\
& & \qquad \qquad \times \;G_1^7(\vu^\alpha,\vu',\vsigma^\alpha)\,
f(\vJ)F(M)\sel(\vu'),
\end{eqnarray}
 were the Jacobian is
simply
 \[
\left|\frac{\p(\vtheta',\vJ',M')}{\p(\vu')}\right| =s^6\cos b
= {\cos b\over\varpi^6} .
\]
Then for each star we choose a sampling density $\xi(\vu'|\vu^\alpha)$
that causes the sample points to be concentrated in the region of
$\vu'$ space that dominates the integral.  In the case of small
errors, this region is the inner $\sim3\sigma$ of the error
ellipsoid. In the case of serious errors -- for example if a variable
such as $\vlos$ has not been measured or the measured value of
$\varpi$ is negative -- this region is a subset of the error ellipsoid
that is consistent with reasonable expectations of how stars are
distributed in phase space.  Again assuming that the measured values
of $l^\alpha,b^\alpha$ and $m^\alpha$ are exact, we are led to the
sampling density
 \[ \label{eq:sampledensdata}
\xi(\vu'|\vu^\alpha) = \left\{
\begin{array}{ll}
0 \qquad \hbox{for }(l,b) \ne (l^\alpha,b^\alpha) \\
 C^\alpha
  {\displaystyle \left|\frac{\p(\vtheta',\vJ',M')}{\p(\vu')}\right|}
  G_4^7(\vu^\alpha,\vu',\vsigma^\alpha) \\
  \qquad\times f_S(\vx',\vv')
  F(m^\alpha,s)\qquad {\rm otherwise,} \\
\end{array}\right.
\]
 where $C^\alpha$ is the normalising constant. With this sampling density the
integral (eq.~\ref{eq:LalphaJx}) reduces to the Monte-Carlo sum
 \[ \label{eq:calcStack}
\like^\alpha_\ast = \frac{\norm}{N\,C^\alpha}\, \sum_{k=1}^N
\frac{f(\vJ'_k)}{f_S(\vx'_k,\vv'_k)} ,
\]
 where $(\vx'_k,\vv'_k)$ is determined from $\vu'_k$, and
$\vJ'_k=\vJ(\vx'_k,\vv'_k)$. We note that $C^\alpha$ is independent of
$f(\vJ)$ and $\Phi$, so it does not vary as we explore different \df s and
potentials. Since we are only ever interested in the ratio between
likelihoods calculated for different \df s and potentials, we do not need to
compute $C^\alpha$.

\subsection{Choice of sampling density}

A good choice of the sampling density $f_S(\vx,\vv)$ used in these
calculations reduces numerical noise by making the individual
contributions to the Monte-Carlo sums as nearly equal as possible. We
achieve this goal by choosing $f_S(\vx,\vv)$ to be a good guess at the
phase-space distribution of the population that our catalogue
samples. It is important to ensure that there are no points with
$f_S\ll f$, as these will then dominate the integrals
(eqs.~\ref{eq:AStack}~\&~\ref{eq:calcStack}), making them very
noisy. We have based these on a product of a double-exponential density in
real space with a triaxial
Gaussian velocity distribution. The principal axes of the velocity
distribution are aligned with the $R,z$ and $\phi$ directions, with
dispersions that vary in proportion to $\exp(-R/8\kpc)$. The means of
the $v_R$ and $v_z$ components are zero, while the mean of $v_\phi$ is
 \[
\langle v_\phi\rangle= v_{\rm c} - v_{\rm a}(R)
\]
with constant $v_{\rm c}=245\kms$, and asymmetric drift velocity
$v_{a}(R)\propto\sigma^2$. 

We use the sum of two such discs (approximating the thin and thick
discs) as $f_S$. We set the dispersions at the Solar radius
$\sigma_{R,\odot} = \sigma_{z,\odot} = 30\kms$, $\sigma_{\phi,\odot} =
40\kms$, and $v_{a,\odot} = 15\kms$ for a disc with scaleheight
$0.3\kpc$; and $\sigma_{R,\odot} = \sigma_{z,\odot} = 50\kms$,
$\sigma_{\phi,\odot} = 60\kms$, and $v_{a,\odot} = 30\kms$ for a disc
with scaleheight $1\kpc$. 30 per cent of $f_S$ is associated with the
thick disc component, and the rest with the thin disc. 

\section{The problem with orbit-based methods}\label{sec:problem}

To understand why orbit-based methods are extremely ill-suited to determining
the likelihood of a stellar catalogue, we need to look at the numerator of
equation~(\ref{eq:calcTorus}). The main calculation for each star $\alpha$ is
finding the line-of-sight integral $\LOSI_{k,\alpha}$ for each orbit used in
the Monte-Carlo integration. These are then summed (with some weights) to
find the probability that this star is predicted by the model.
$\LOSI_{k,\alpha}$ is essentially the integral down the line-of-sight of the
probability of a given orbit giving the observed value of $\vu$. With increasingly
precise data, the number of orbits for which this probability is
non-negligible anywhere along the line of sight diminishes. 

The computation of a star's likelihood proceeds by considering each point
along the line of sight to the star, and then for each orbit finding the
velocities that a star on that orbit will have at that point.
Each such velocity then contributes to the probability a factor proportional
to 
 \[
\exp\left[-\left({(\vlos^\alpha-\vlos)^2\over2\sigma_\parallel^2}
+{|\vmu^\alpha-\vmu|^2\over2\sigma_\mu^2}\right)\right].
\]
 As we proceed along the line of sight, the observables $(\vlos,\vmu)$
predicted by the orbit gradually change, so the overall contribution of the
orbit to the likelihood comes from a line drawn through the error
ellipsoid. The contribution is large or small depending on how close the
line comes to the centre of the ellipsoid.

When we change potential, our orbits necessarily change and the lines through
the error ellipsoid change. It may happen that our new orbit library has an
orbit that yields a line that comes very close the centre of the ellipsoid,
whereas the old orbit library jumped from an orbit that gave a line passing
closest to the centre say $2\sigma$ on one side to an orbit that passes
closest $2\sigma$ on the other side of the ellipsoid's centre. This old library
provided no orbit that makes the given star very probable even though, with a
denser sampling of action space, such an orbit would have arisen.  Hence with
the old library rather than the new, the star is declared improbable even
though it is in fact probable.

A subsidiary issue is that with the new library some of the values $\vJ_k$
which had $\phi(\vJ_k)=0$ in the original potential (i.e. orbits that do not
enter the survey volume, and were thus irrelevant to our calculation), have
$\phi(\vJ_k)\neq0$ in the new library and have to be considered, and vice
versa. 

An indication of how significant these discreteness effects are is that
changing the potential from one with scaleheight $300\pc$ to one with
scaleheight $350\pc$ causes the calculated values $\LOSI_{k,\alpha}$ to
change by on average a factor $\sim 10$.

\subsection{Extent of the problem}\label{sec:extent}

To see the impact of this problem, we now turn to our three catalogues
described in Section~\ref{sec:models}. Numerical experiments show that
keeping the same values of $\vJ_k$ as we change potentials is little better
than taking an entirely new Monte-Carlo sampling of $\vJ_k$, so we now explore
the case in which we keep the same Mc11 potential, $\Phi$, and the same
catalogue of stars, but use a new sample of actions $\vJ_k$ to calculate the
likelihoods (eq.~\ref{eq:calcTorus}). In each case we use the \df\ $f_{\rm
true}$ that was used to create the catalogue as both the sampling density and
the \df\ being tested in equation~(\ref{eq:calcTorus}). 

In principle it would be ideal to take a large number of Monte-Carlo samples,
each containing $N_k$ values of $\vJ$, determine the likelihood in each, and
directly determine the scatter. However this process is prohibitively
expensive computationally, so we use a quicker alternative. We calculate the
integrals $\phi_k$ and $\mathrm{LOSI}_{k,\alpha}$ (equations~\ref{eq:phi}
\& \ref{eq:LOSI} respectively) for $100\,000$ values of $\vJ_k$. Then we
choose at random subsets of $12\,500$ or $25\,000$ or $50\,000$ values of
$\vJ_k$ and use these values to calculate for each star $\alpha$ the standard
deviation $\sigma(\log\mathcal{L}_\ast^\alpha)$ of the resulting values of
$\log\mathcal{L}_\ast^\alpha$. 

\begin{figure*}
\centerline{
\includegraphics[width=.3\hsize]{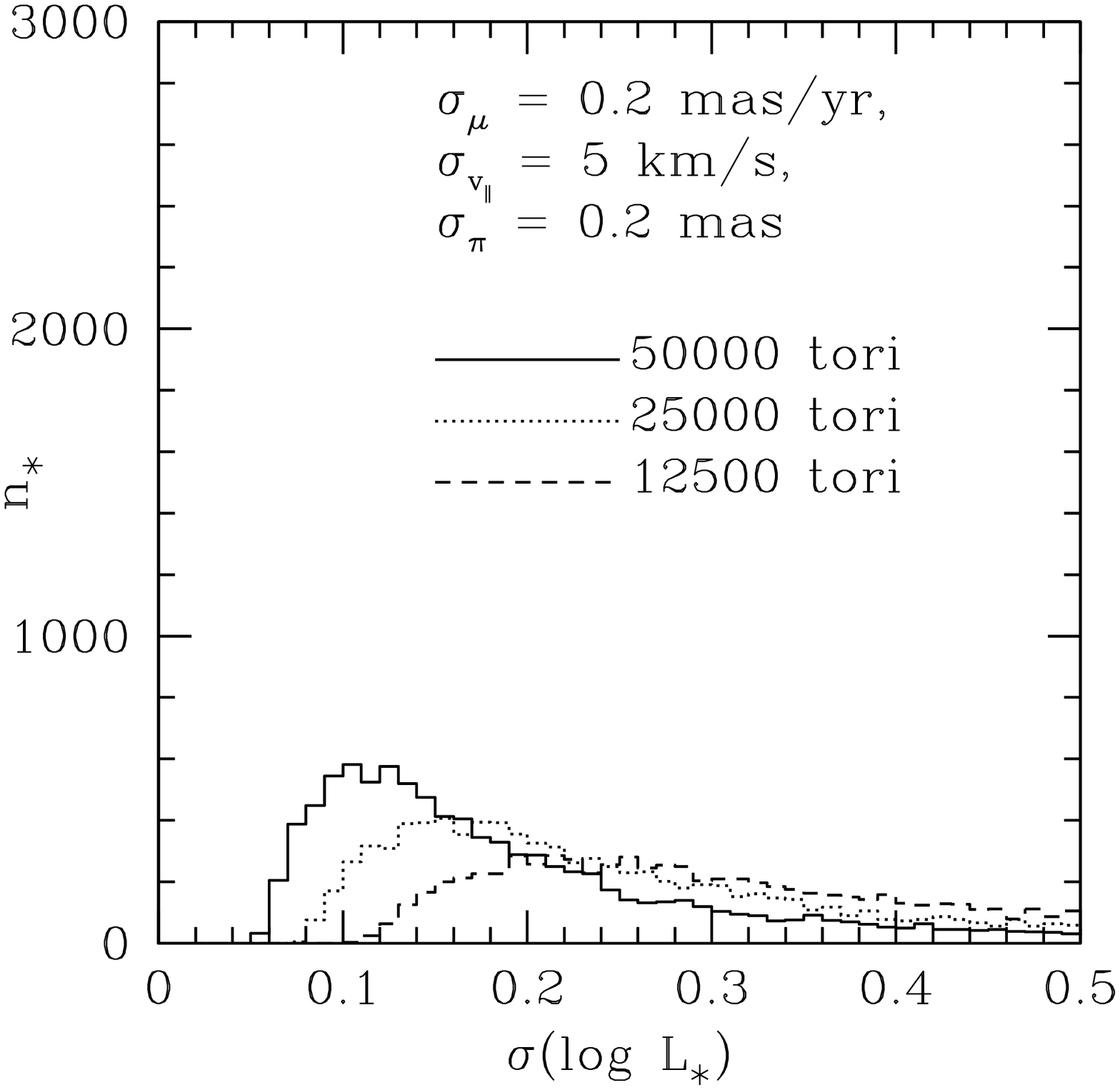}
\includegraphics[width=.3\hsize]{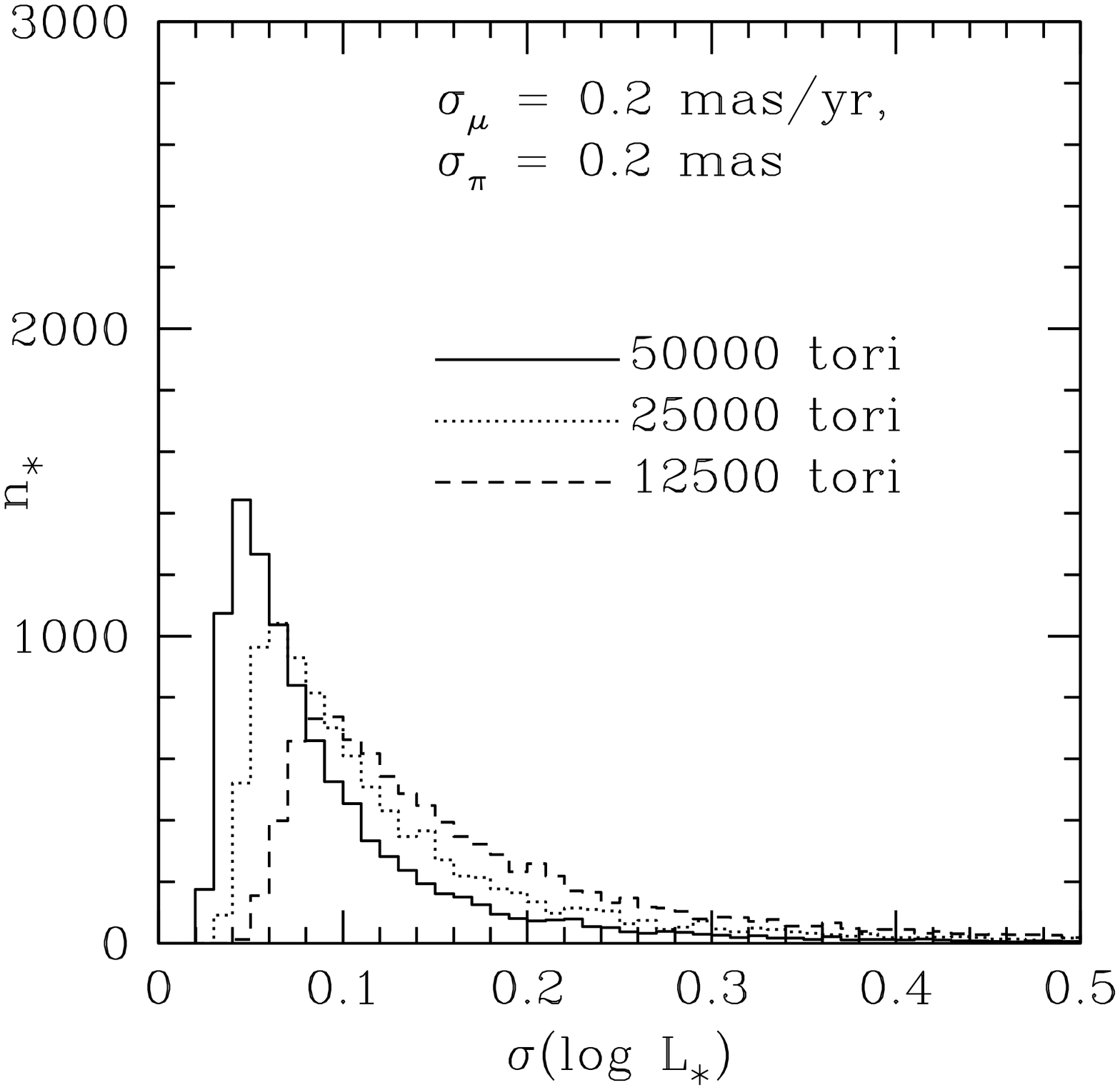}
\includegraphics[width=.3\hsize]{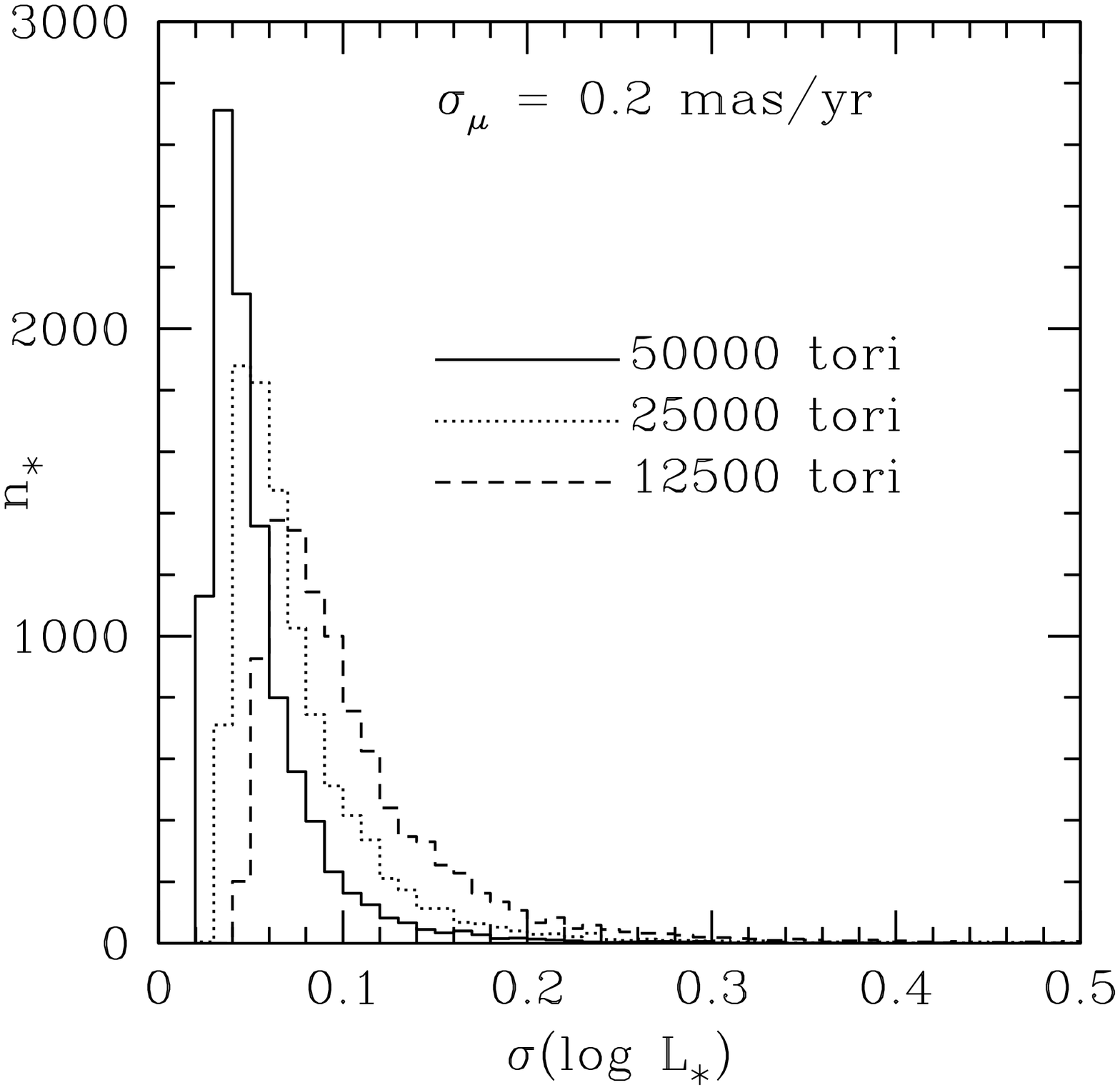}
}
 \caption{Torus modelling: histograms of the standard deviations
$\sigma(\log\like_\ast)$ calculated from torus libraries of various sizes (as
labelled) for each of the $10\,000$ stars in each of  our three catalogues (also as
labelled).} \label{fig:dlstar}
\end{figure*}

Fig.~\ref{fig:dlstar} shows the distributions of these standard deviations
for each library size and each quality of data.  Increasing the quality of
the data broadens the distribution of $\sigma(\log\mathcal{L}_\ast^\alpha)$
and shifts its mean to higher values.  Even for $N_k=50\,000$ all stars have
non-negligible values of $\sigma(\log\mathcal{L}_\ast^\alpha)$ and only in
the case of the highest-quality data do the distributions have a heavy tail
to large values. Thus the problem is not so much the existence of a few
outliers but that our Monte-Carlo sums give rise to excessive uncertainty in
the likelihoods of all stars. 

Note that the largest torus libraries considered are
more than an order of magnitude larger than the orbit libraries used for
typical Schwarzschild modelling of external galaxies
\citep[e.g.][]{vdBea08}, though we do not ``dither'' orbits, which is an
approach that can significantly increase the effective resolution of
Schwarzschild models.

\begin{figure}
  \centerline{\resizebox{\hsize}{!}{\includegraphics{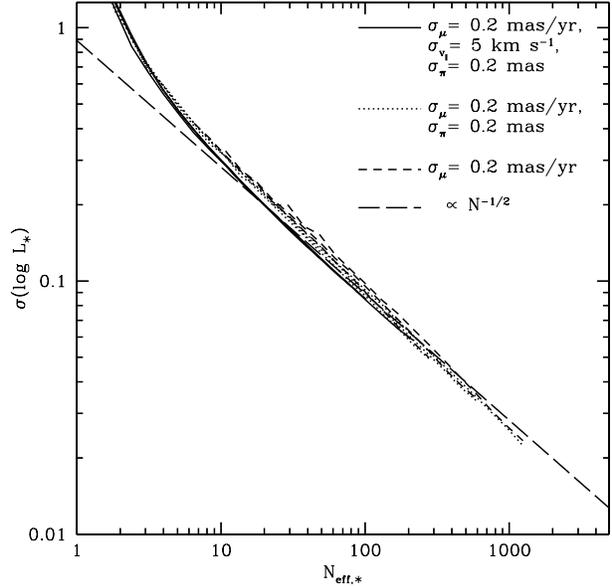}}}
  \caption{Torus modelling: standard deviations of a stars' likelihood, 
$\sigma(\like_\ast)$ as a function of the
effective number of tori contributing to the calculation
of each likelihood ($N_{\mathrm{eff,\ast}}$
eq.~\ref{eq:neff}). The three star catalogues, which
differ in the completeness of their data, are
represented by solid, short-dashed or dotted lines. There are
three lines for each catalogue in the figure, corresponding to
three different sizes of torus library, $N_k=12500, 25000$ and
$50000$. The long-dashed line is $\sigma(\like_\ast)=0.9\,
N_{\mathrm{eff,\ast}}^{-1/2}$, which fits all points with
$N_{\mathrm{eff,\ast}}\gtrsim 10$ quite well. }
 \label{fig:dll_vs_Ne}
\end{figure}

From Fig.~\ref{fig:dlstar} it is evident that
$\sigma(\log\mathcal{L}_\ast^\alpha)$ can be beaten down by increasing the
number $N_k$ of orbits in one's library.  In Paper I we used the Shannon
entropy to measured the extent to which an individual observation
$\vu^\alpha$ is probed by a given library of tori. The entropy is
 \begin{equation}
S^\alpha=-\sum_k^N p^\alpha_k\ln p^\alpha_k,
\end{equation}
 where $p^\alpha_k$ is the fraction of the calculated likelihood
$\like^\alpha_\ast$ contributed by the $k$th torus. This is
 \begin{equation}\label{eq:defspk}
p^\alpha_k \equiv
\frac{\LOSI_{k,\alpha}}{\sum_k\LOSI_{k,\alpha}} .
\end{equation}
 Clearly $S^\alpha=0$ if there is only one contribution, and $S^\alpha=\ln
N$ if $N$ tori provide equal contributions to the star's probability. We
can therefore define an effective number of contributing values $\vJ_k$,
 \begin{equation} \label{eq:neff}
N^\alpha_{\mathrm{eff,\ast}} \equiv \exp(S^\alpha),
\end{equation}
 which is the number of tori with equal $p^\alpha_{k}$ that would give an
entropy $S^\alpha$. Clearly we expect that, for a given library size $N_k$,
the typical value of $N^\alpha_{\mathrm{eff,\ast}}$ will become smaller as
the observational data become more precise.

Figure~\ref{fig:dll_vs_Ne} plots the standard deviation
$\sigma(\log\mathcal{L}_\ast^\alpha)$ against the effective number
$N_{\mathrm{eff,\ast}}^\alpha$ derived from the average of the entropies
$S^\alpha$ over a sample of equivalent torus libraries. We see that all data
points lie close to a universal relation, that is independent of library size
or data quality. Moreover, this relation asymptotes to the relation
$\sigma(\log\mathcal{L}_\ast^\alpha)\propto\surd
N_{\mathrm{eff,\ast}}^\alpha$.  This result conclusively proves that the
scatter in likelihood values is generated by Poisson noise, and enables us to
predict how large $N_{\mathrm{eff,\ast}}^\alpha$ needs to be to beat the
noise down to any given level for any data set.

\begin{figure*}
\centerline{
\includegraphics[width=.3\hsize]{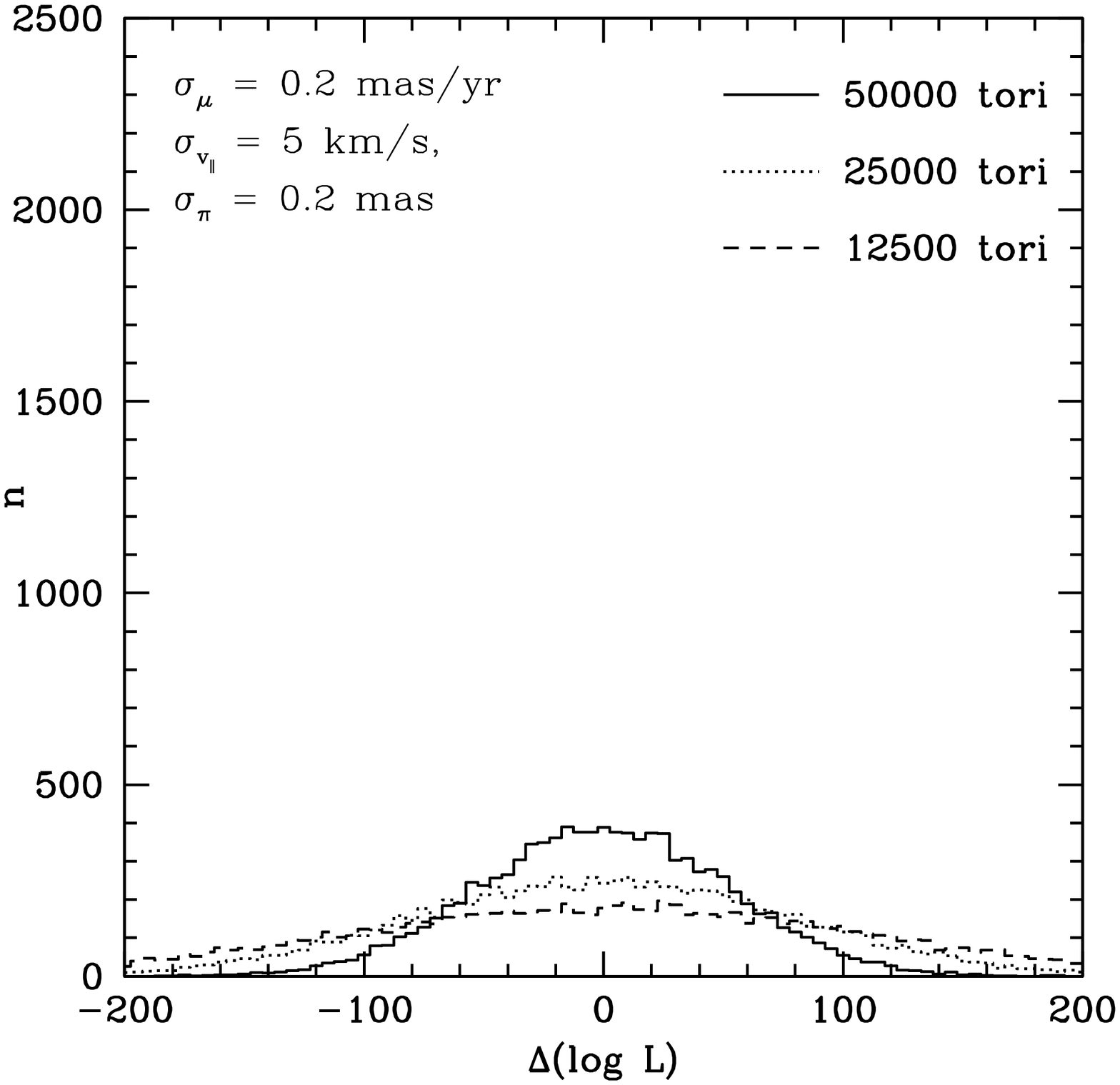}
\includegraphics[width=.3\hsize]{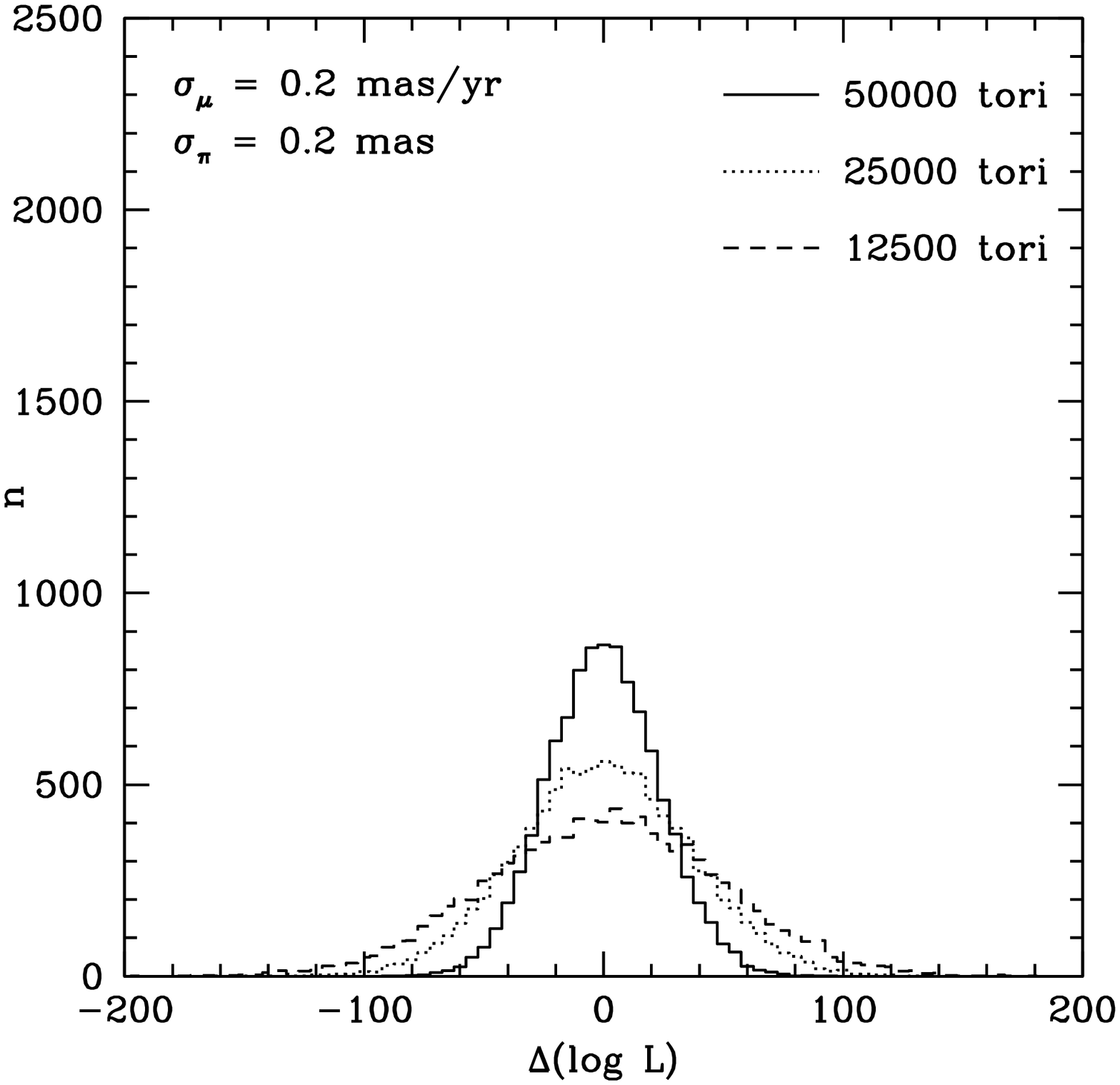}
\includegraphics[width=.3\hsize]{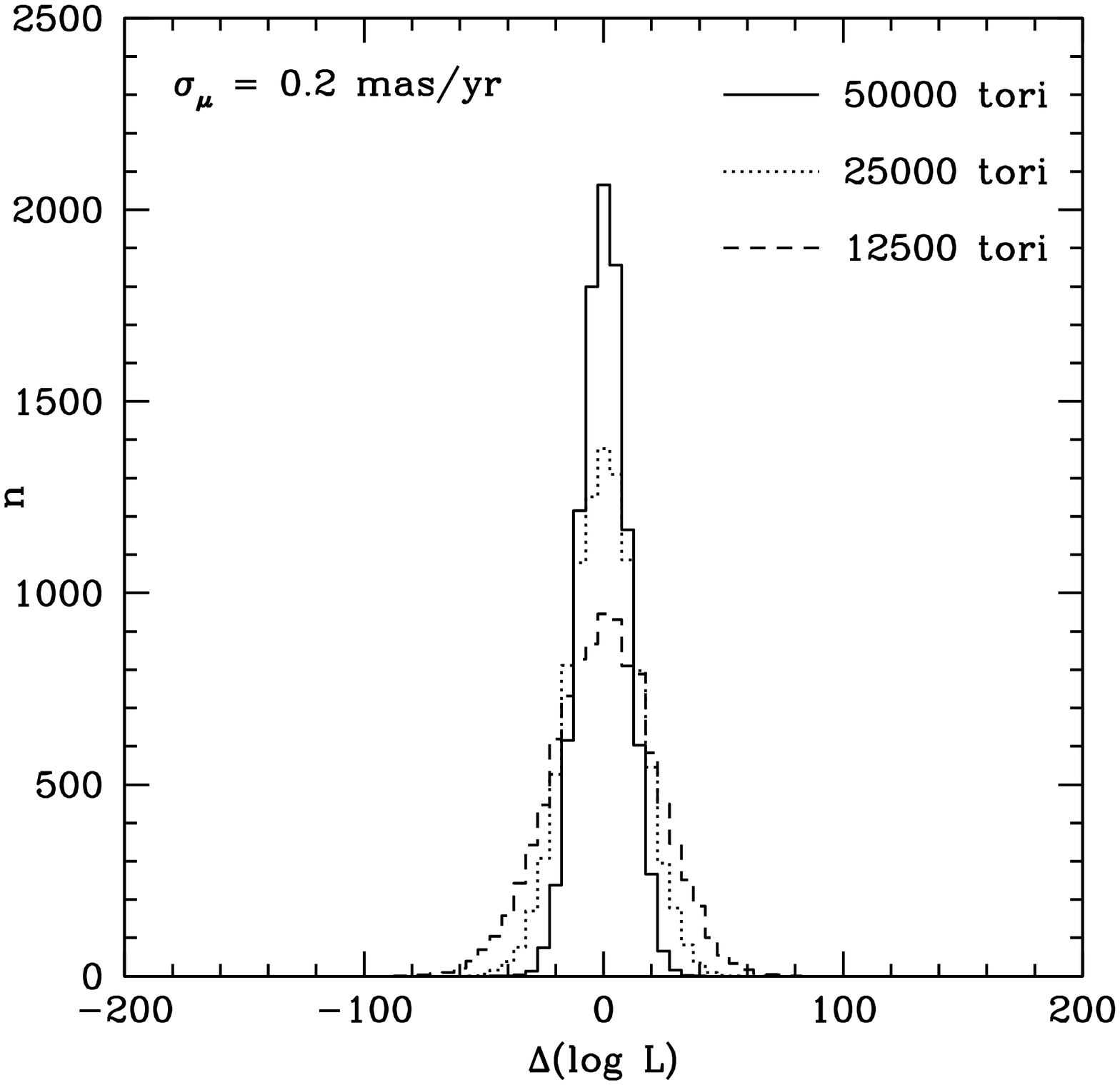}
}
 \caption{Torus modelling: histograms of the value of $\log\like$ calculated for
each of our three catalogues. Each value of $\log\like$ is computed $10\,000$
times using a different torus library. Distributions are shown for three
sizes of torus library: $N_k=12\,500, 25\,000$ or $50\,000$ tori. From each
value of $\log\like$ we have subtracted the mean value of its set. 
 } \label{fig:dlmod}
\end{figure*}

Figure~\ref{fig:dlmod} shows how the errors in individual star likelihoods
combine to produce errors in model likelihoods $\like$ by showing histograms
of the offsets $\Delta(\log\like)$ between $\log\like$ and its mean over all
torus libraries. These distributions are roughly Gaussian so their widths are
characterised by the standard deviation $\sigma(\log\like)$ of $\log\like$.

Figure~\ref{fig:universal} shows $\sigma(\log\like)$ as a function of torus
library size $N_k$ for all three catalogues. The more precise the data, the
larger $\sigma(\log\like)$ is, but in each case $\sigma(\log\like)$ declines
approximately in proportion to $\sqrt{N_k}$, but is still $\sim10$ even for
our largest library and our lowest quality data.

\subsubsection{Implications for determination of $\Phi$}

\begin{figure}
\centerline{\includegraphics[width=.8\hsize]{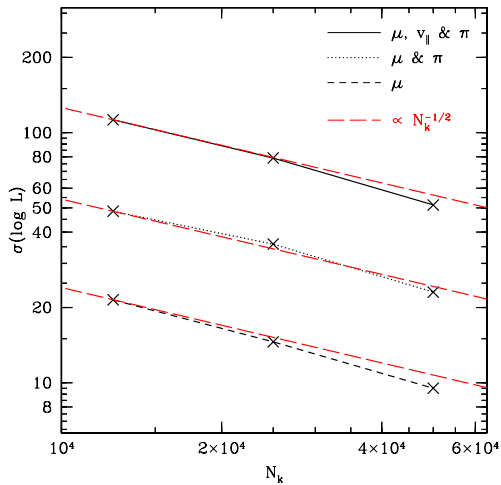}
}
 \caption{Torus modelling: the standard deviations of the histograms plotted
in Fig.~\ref{fig:dlmod} as functions of the number $N_k$ of tori employed.
The red long-dashed lines have slope $-1/2$ so we see that the
uncertainty in $\log\like$ declines approximately as
$N_k^{1/2}$.}\label{fig:universal}
\end{figure}

\begin{figure*}
\centerline{
\includegraphics[width=.3\hsize]{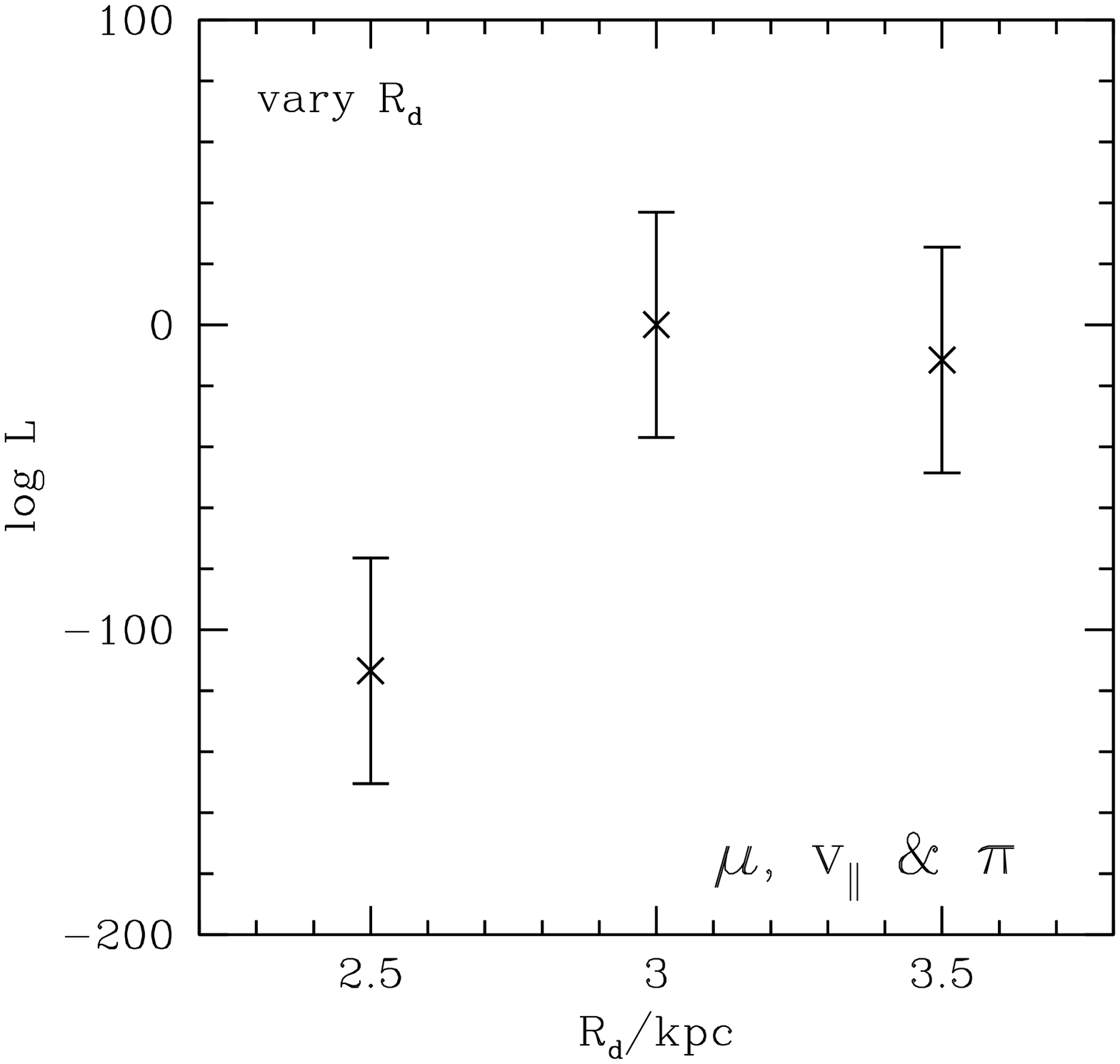}
\includegraphics[width=.3\hsize]{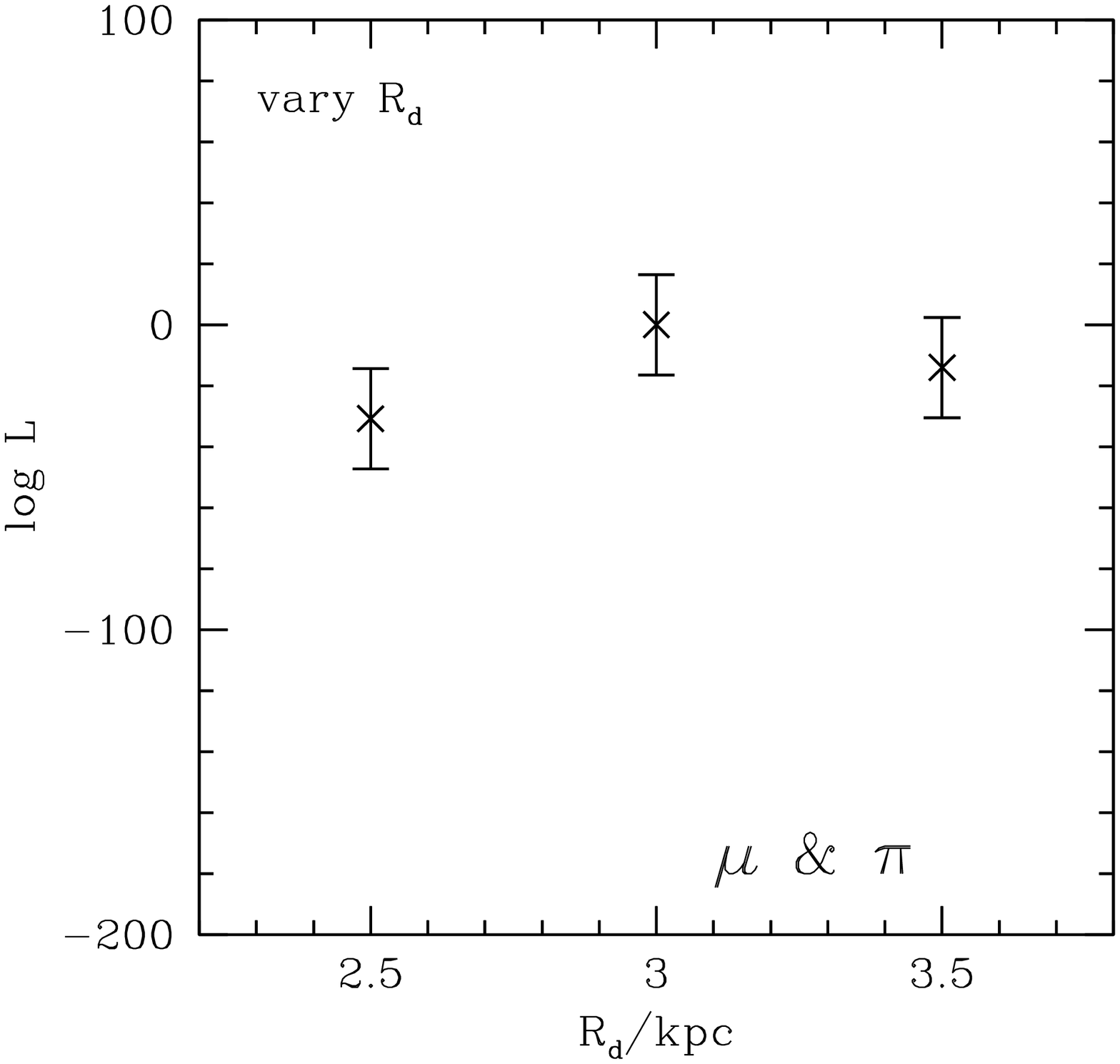}
\includegraphics[width=.3\hsize]{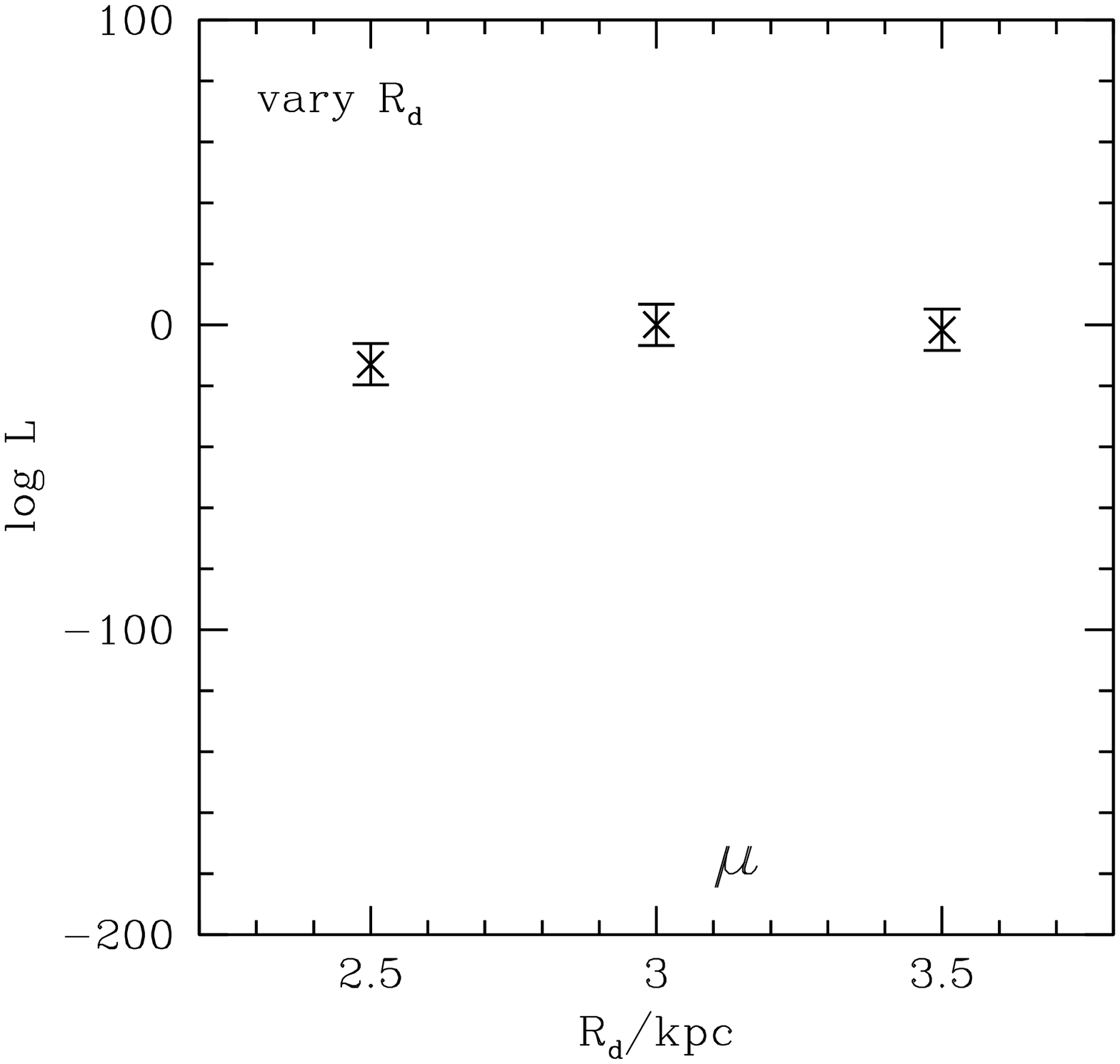}
}
\centerline{
\includegraphics[width=.3\hsize]{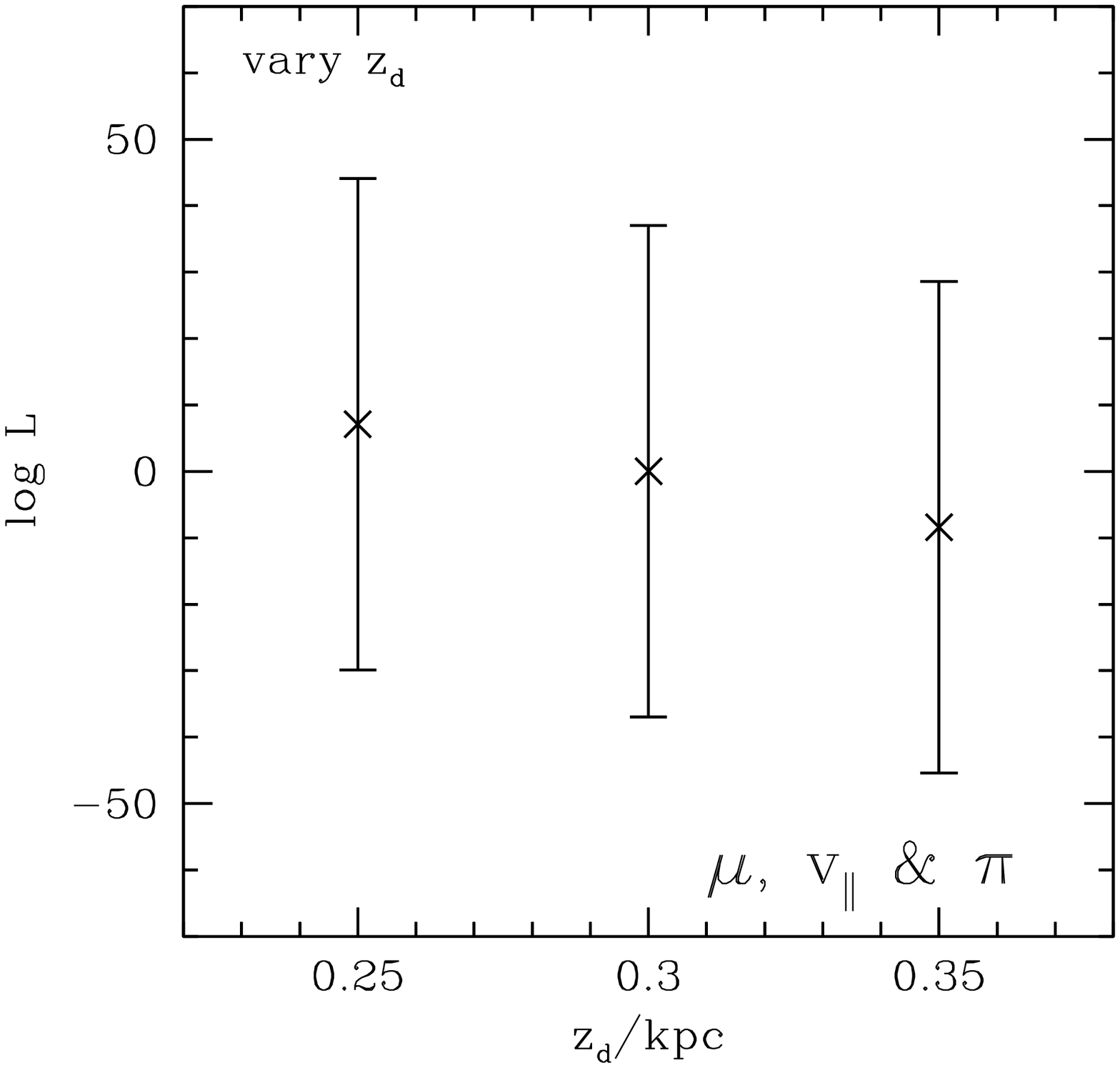}
\includegraphics[width=.3\hsize]{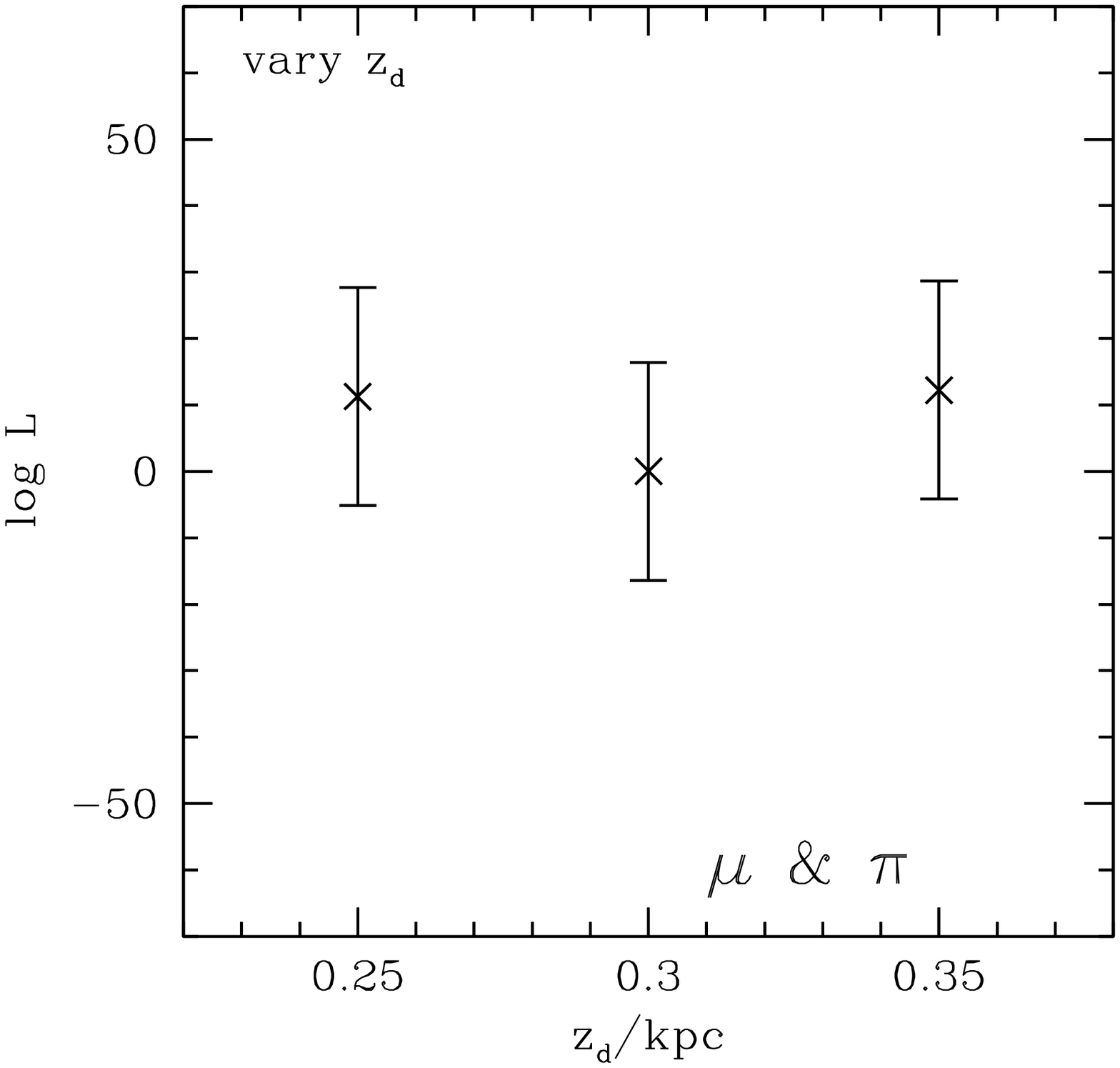}
\includegraphics[width=.3\hsize]{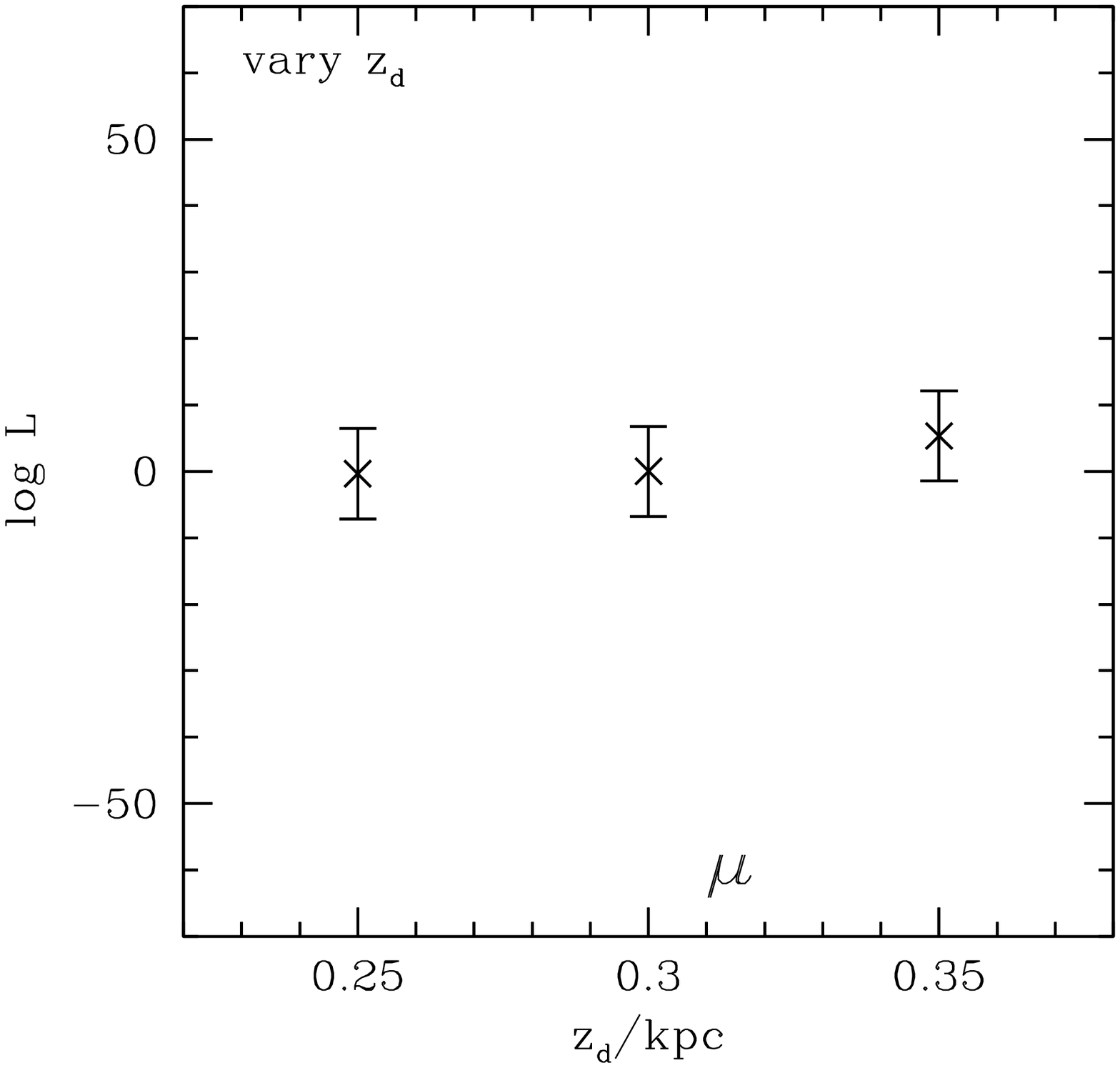}
}
 \caption{Torus modelling: differences in the largest values of log
likelihood obtained by varying the \df\ in a given $\Phi$ and the value for
the true $\Phi$. The \df\ $f(\vJ)$ is assumed to be of the form
(\ref{totalDF}). In the upper row the scalelength $R_\d$ of the disc that
contributes to $\Phi$ is varied, while in the lower row the its scaleheight
$z_\d$ is varied. The true values are $(R_\d,z_\d)=(3,0.3)\kpc$. The
computations use $N_k=100\,000$ tori. The error bars are approximate, and
found from an extrapolation to $N_k=100\,000$ of the relations shown in
Fig.~\ref{fig:universal}, assuming that $\sigma \log\like \propto
N_k^{-1/2}$.} \label{fig:llpots_tori}
\end{figure*}

The key question is: will this uncertainty on the value of $\like$ prevent us
from making useful inferences about the Galactic potential? To answer this
question we now use tori to determine the likelihoods of our three catalogues
given various choices of potential. In Figure~\ref{fig:llpots_tori} we show
the peak values of $\log\like$ calculated for models constrained to have two
pseudo-isothermal discs in potentials with varying disc scalelengths (upper
row) and scaleheights (lower row) found from  a Monte-Carlo sum over
$100\,000$ values $\vJ_k$ (i.e.\ $100\,000$ tori). The true potential has
disc parameters $(R_\d,z_\d)=(3,0.3)\kpc$.
We show (approximate) error
bars on each value -- these are found by extrapolating the relationship shown in
 Figure~\ref{fig:universal} to $N_k=100\,000$. Note,
therefore, that this is the uncertainty on $\like$ due to the imperfect
analysis, and is not intrinsic to the observational data.

In each case, the models with $R_\d=2.5$ can clearly be ruled out, but the
models with $R_\d=3.5$ have the same $\like$ as the true potential to within
the error bars. The results when varying $z_\d$ are even less encouraging, as
all the calculated values of $\like$ agree to within the error bars, and the
incorrect potentials ($z_\d=0.25$ or $0.35$) are sometimes preferred to the true
potential. Clearly the true difference in $\log\like$ is significantly smaller
than the uncertainty. 

These calculations would require over a week on a single desktop cpu
for each potential (though they are simple to parallelise). The
majority of this time is spent calculating the line-of-sight integrals
$\mathrm{LOSI}_{k,\alpha}$.

\begin{figure}
\centerline{
\includegraphics[width=.5\hsize]{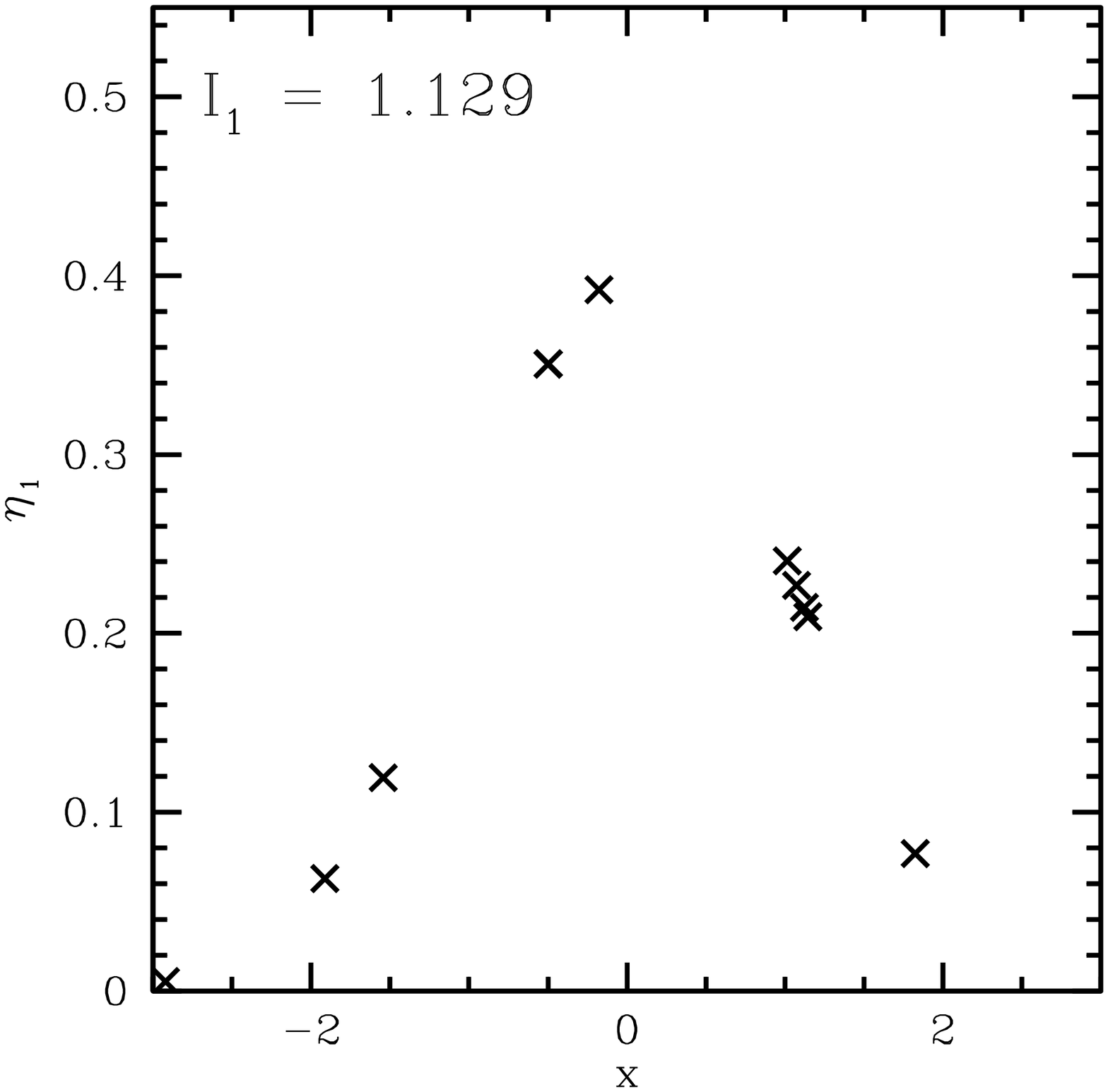}
\includegraphics[width=.5\hsize]{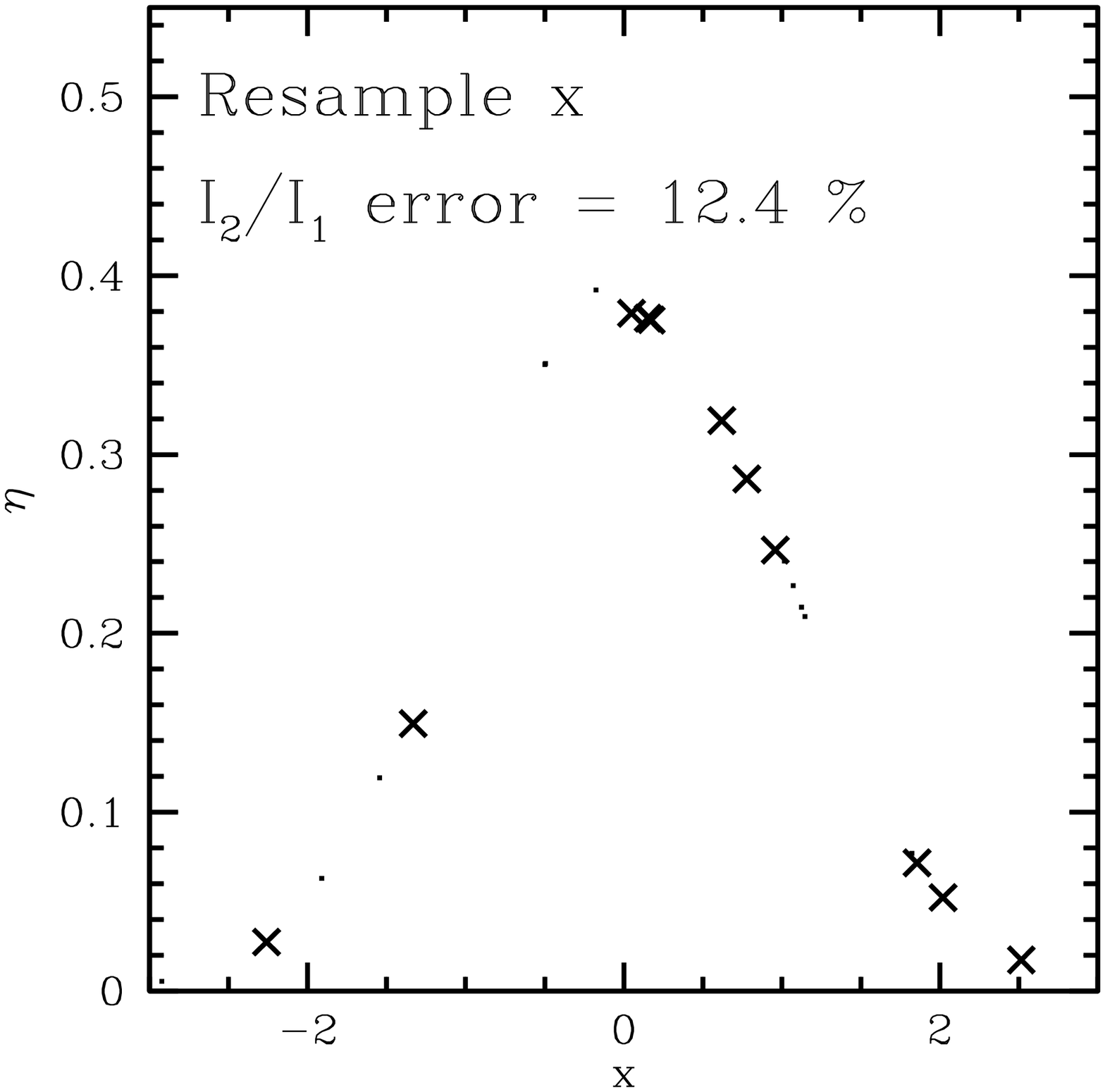}
}\centerline{
\includegraphics[width=.5\hsize]{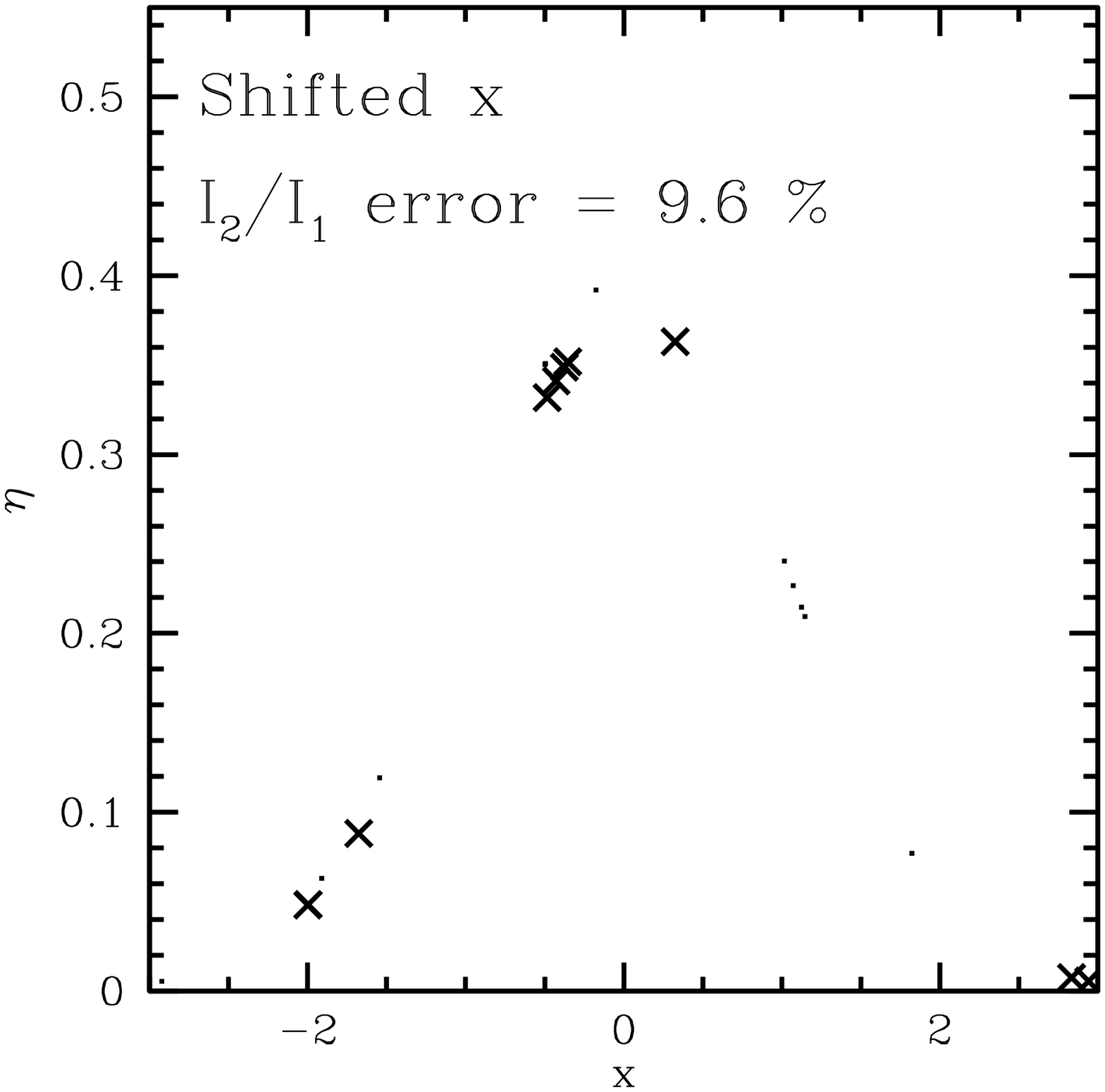}
\includegraphics[width=.5\hsize]{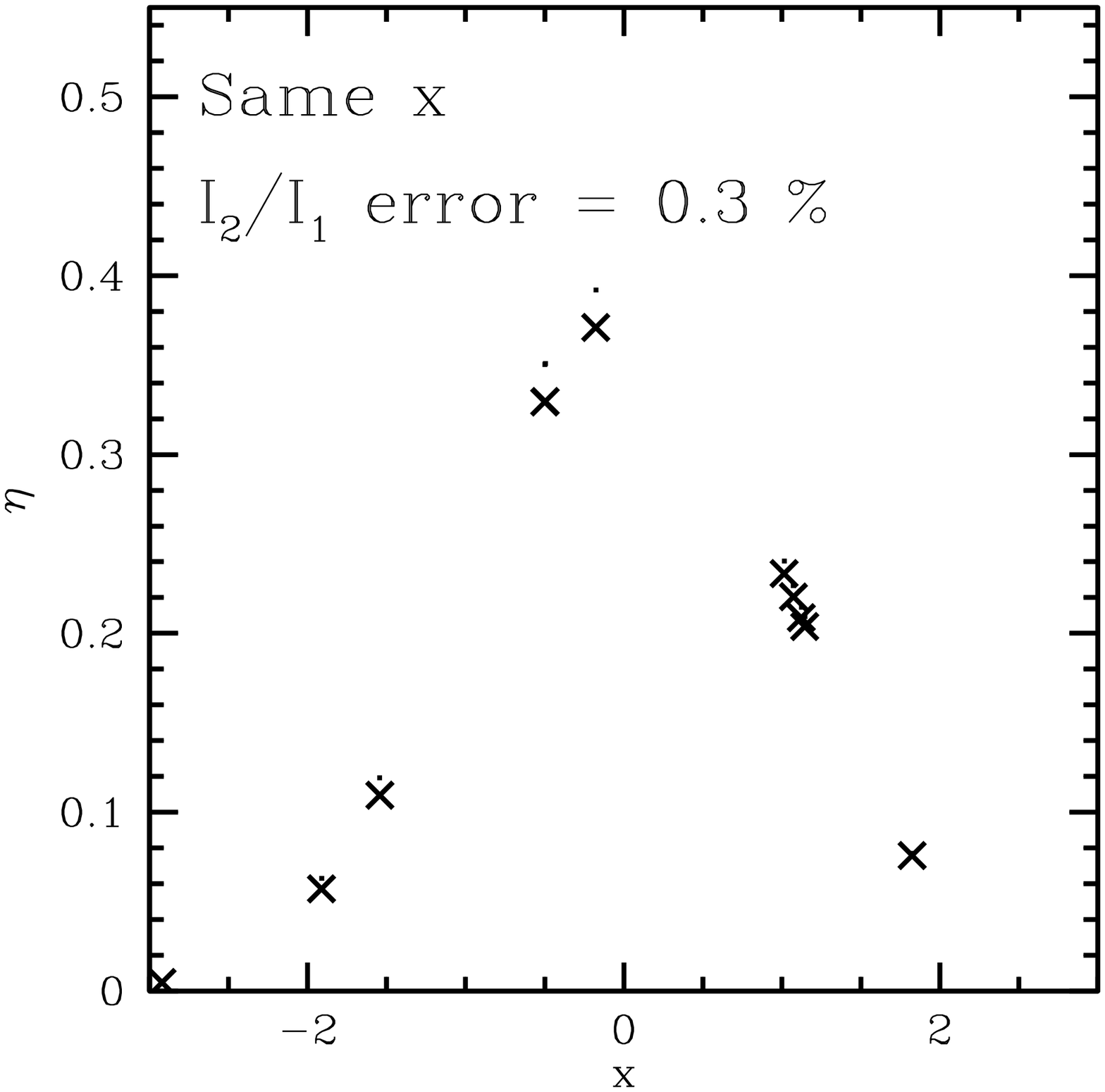}
}
 \caption{Toy example analogous to comparing the likelihoods of given data in
different potentials. In each case we determine an integral $I_i$
(eq.~\ref{eq:I}) by Monte-Carlo summation over $20$ points taken randomly
from the range $-5<x<5$. In the top-left panel we show the function values
$\eta_1$ used to find $I_1$ (in the other panel these values are shown as
dots, for comparison). The sum yields $I_1=1.129$ rather than unity. In the
other three panels we show the summed values of $\eta_2$ when we have either
completely resampled the values of $x$ (\emph{top right}), or have shifted
all the values by $\Delta_x=-1.5$ (with any points that fall below $x=-5$
replaced by ones at $x>3.5$, \emph{bottom left}), or have used the same
values of $x$ as were used for $I_1$ (\emph{bottom right}). Although these
last points do not yield a particularly accurate value of $I_2$ they do
yield an accurate value for the ratio $I_1/I_2$.} \label{fig:toy}
\end{figure}

\begin{figure}
\centerline{
\includegraphics[width=.48\hsize]{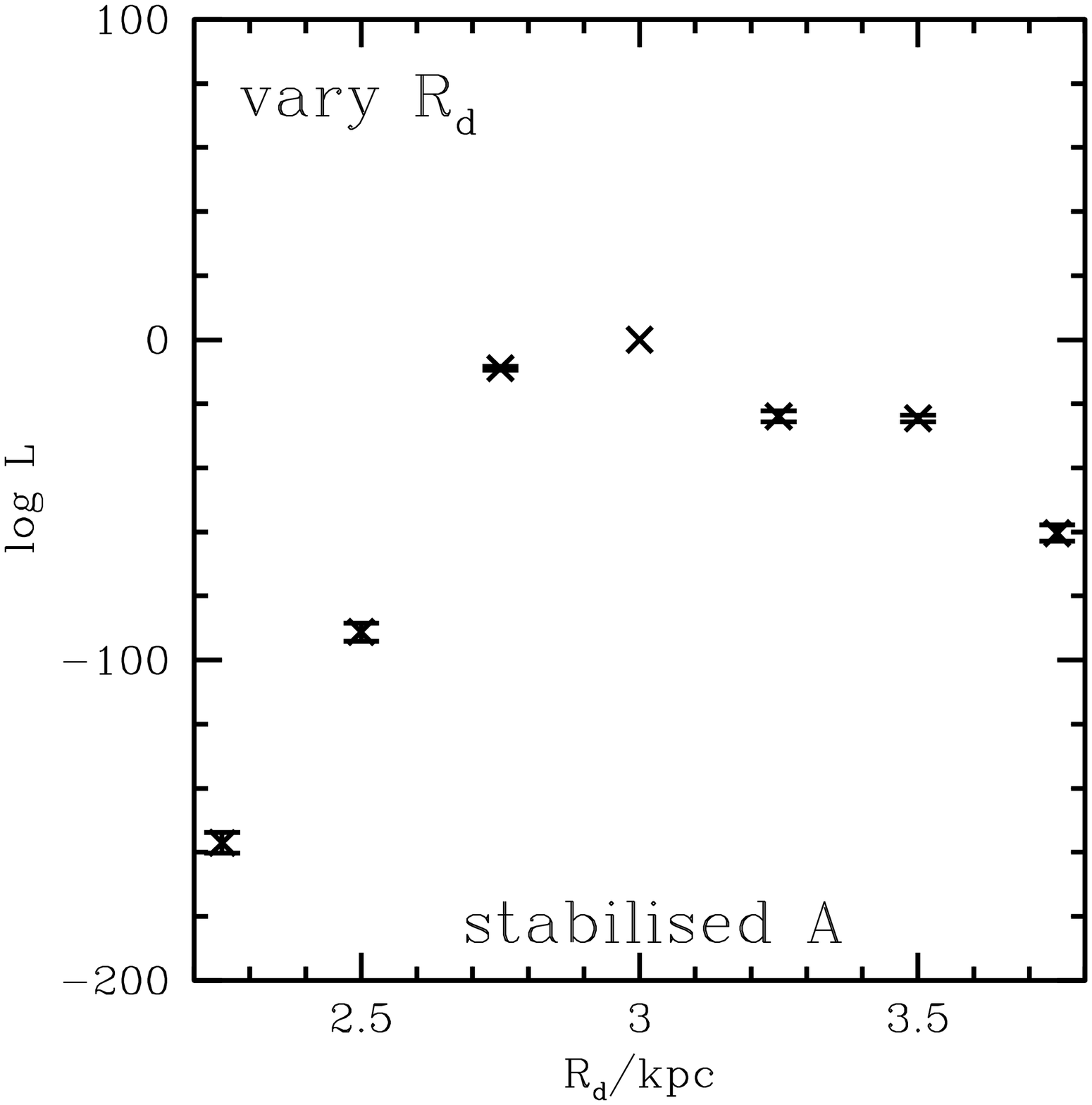}
\includegraphics[width=.48\hsize]{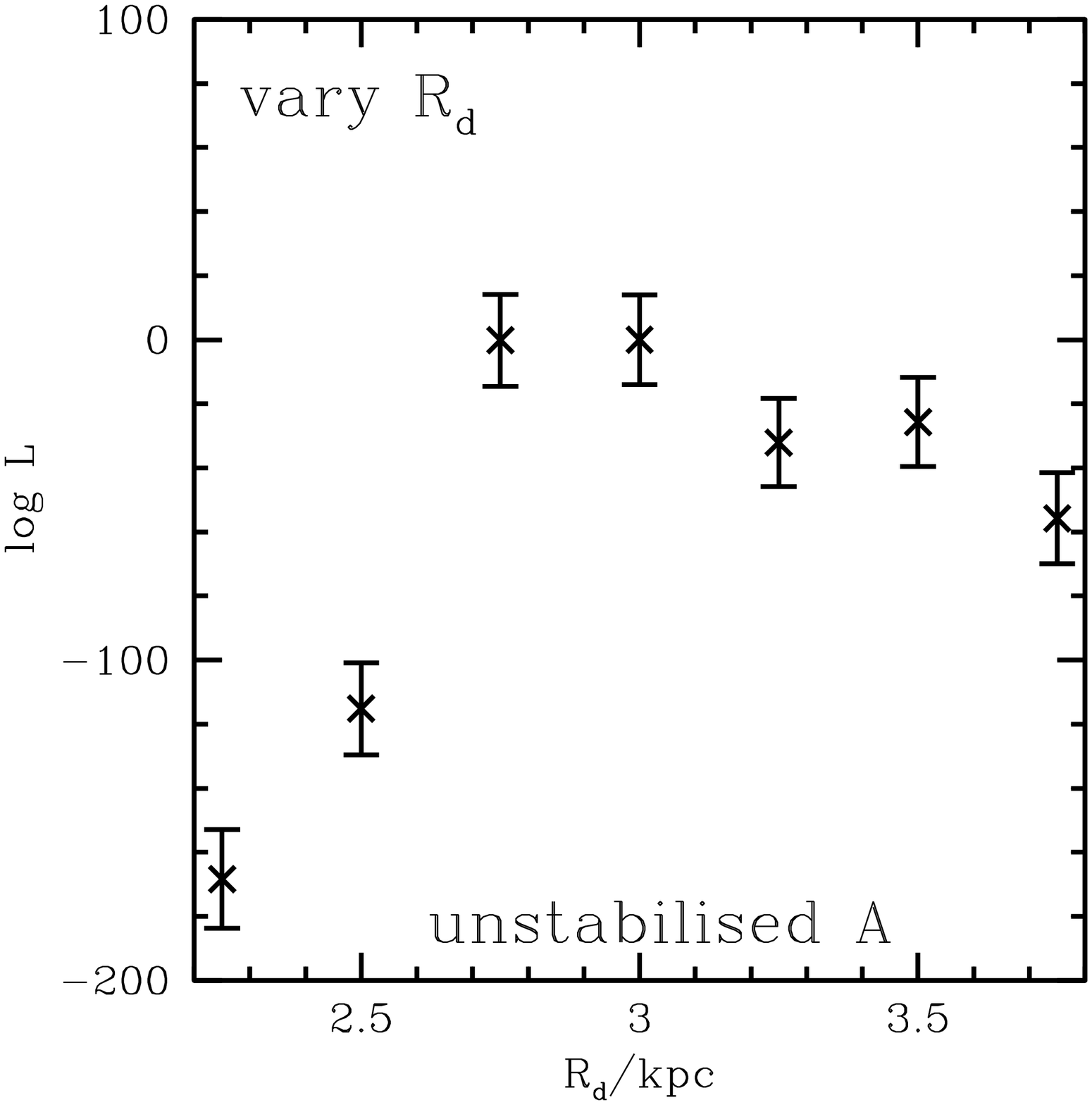}
}\centerline{
\includegraphics[width=.48\hsize]{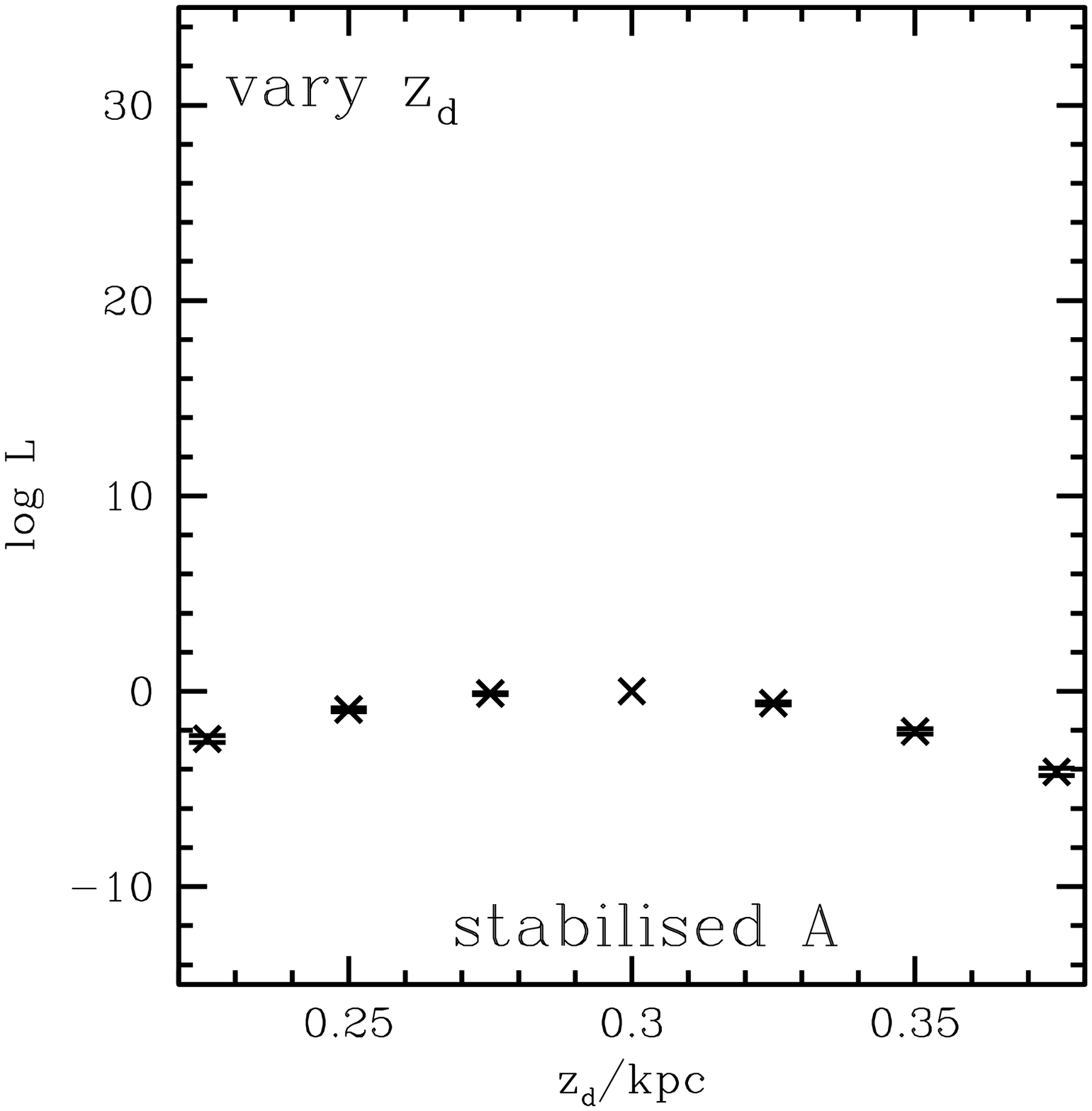}
\includegraphics[width=.48\hsize]{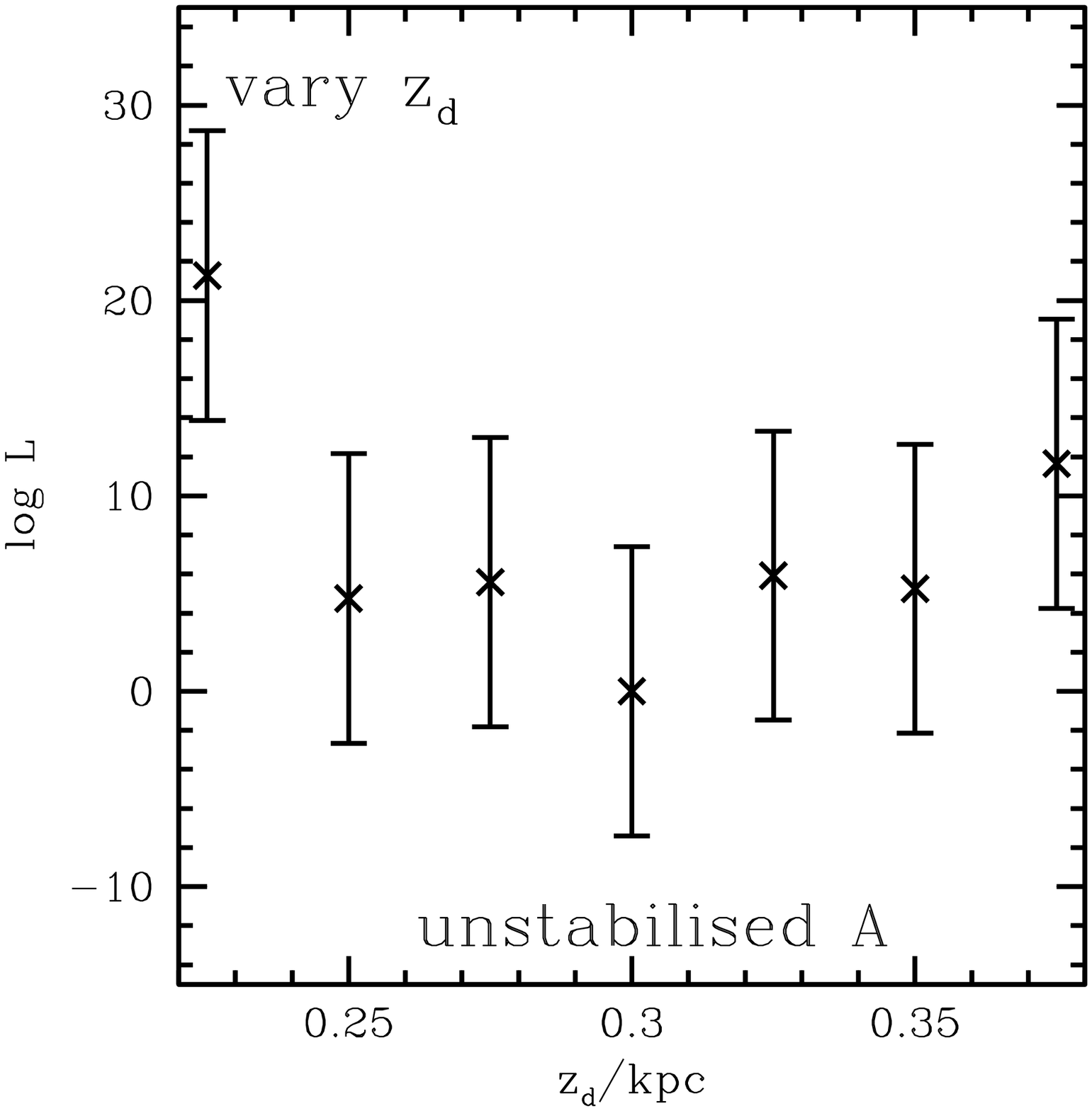}
}
 \caption{Using $\vJ(\vx,\vv)$ when the data are error-free: Differences
between the largest value of $\log\like$ obtained for a candidate $\Phi$ and
the value obtained for the true $\Phi$ as we vary either the scalelength of
the potential-generating disc (upper) or its scaleheight (lower). The left
panels show results obtained using the fixed sets of sampling phase-space
points throughout, while the right column shows the effect of choosing new
points for the estimation of $\norm$ for each trial $\Phi$.
Note that the range of $\log\like$ on the y-axis is less than half
that in Fig.~\ref{fig:llpots_tori}.}
\label{fig:llpots_st}
\end{figure}

\subsection{Why could we  determine the DF?} \label{sec:toy}

Our aim in this section is to explain clearly why the torus-based method of
Paper I could determine the \df\ to good precision but fails here when
extended to determination of $\Phi$. To this end we consider the toy problem
illustrated in Figure~\ref{fig:toy}.  We use the Monte-Carlo
principle to estimate the ratio of two integrals $I_i$ of functions
$\eta_i(x)$ of one variable that we know to be products of the Gaussian of
unit dispersion and zero mean and functions $f_i(x)$ that vary on scales
larger than unity. Thus
 \[\label{eq:I}
I_i = \int \eta_i(x)\d x = \int\d x\,, f_i(x) \times G(x,0,1),
\]
 where for the $f_i$ we adopt
\begin{eqnarray}
f_1(x)&=&1+0.01\,x\nonumber\\
f_2(x)&=&0.95+0.03\,x.
\end{eqnarray}
 With these choices the true values are $I_1=1$ and $I_2=0.95$.
The Gaussian represents a star's error ellipsoid and the $f_i$ represent
candidate \df s.  Figure~\ref{fig:toy} shows the process of Monte-Carlo
evaluation of the $I_i$. In the top left panel $I_1$ is found to be
$I_1=1.129$ rather than its true value, unity. In the top right panel
independently sampled points are used to estimate that $I_2=1.206$, so the ratio
of the integrals is $I_2/I_1=1.068$ rather than $0.95$ as it should be. The
lower panels show re-evaluations of $I_2$ using points that are not independent
of those used to evaluate $I_1$: in the bottom-right panel we use exactly the
same points to find $I_2/I_1=0.953$, within 0.4\% of its true value, 
while in the bottom left panel we use points that are shifted by $1.5$ to the
left, and find $I_1/I_2=0.860$. 

This experiment shows that if we use the same sampling points we can
determine the ratio of two integrals like those we have to evaluate to obtain
$\like_\ast^\alpha$ much more accurately than we can determine either
integral individually because the Poisson noise in the evaluation largely
cancels from the ratio. If we use different sampling points to evaluate each
integral, the Poisson noise does not cancel and the ratio is even less
accurate than the individual integrals. In Paper I we used one set of
sampling points to evaluate the likelihood of every \df, so the Poisson noise
made little contribution to the differences of the log likelihoods of the \df s
considered, and we could identify the true \df\ accurately. In Section
\ref{sec:extent} we were obliged to vary the sampling points between
potentials and the Poisson noise in the differences of log likelihoods
degraded performance to an unacceptable extent.

\section{The solution: use $\vJ(\vx,\vv)$}\label{sec:solution}

In Section \ref{sec:EJxv} we explained how, if we have a way to find
$\vJ(\vx,\vv)$, the $\like_\ast^\alpha$ can be evaluated using samples of
points $\vu_k'$ that are chosen specifically for each star in the catalogue
and are never varied. We also explained that a second sample of points should
be used to evaluate the normalising constant $\norm$ for all potentials
considered. We now show that this approach dramatically reduces the numerical
noise that was so prominent in Section~\ref{sec:extent}.

 In summary, the scheme is:

\begin{enumerate}

\item For the calculation of $\norm$, sample $N_\norm$ points $(\vx,\vv)$
from the sampling density $f_S(\vx,\vv)$ (ignoring any that lie outside the
survey volume).

\item Sample $N_{\vu'}$ points for each star from the
sampling density $\xi(\vu'|\vu^\alpha)$, (eq.~\ref{eq:sampledensdata}).

\item \label{enum:restart} Choose a gravitational potential $\Phi$,
and determine $\vJ$ for each of the $N_\norm$ points used to
determine $\norm$ and each of the $N_\alpha \times N_{\vu'}$ points
used to then determine $\like$.

\item Maximise $\like$ in this potential by varying the parameters of
the \df\ $f(\vJ)$.

\item Return to step~\ref{enum:restart}, choosing a new potential.

\end{enumerate}

This process is orders of magnitude faster than the torus approach, so
we are able to carry out many more tests.

As a proof of principle, we show in the left column of
Fig.~\ref{fig:llpots_st} results for the case in which we have perfect
observational data with the consequence that the Monte-Carlo sum for
$\like^\alpha_\ast$ (eq.~\ref{eq:sampledensdata}) requires just one
point.  These tests are very similar to those of \cite{YSTea13},
except we have significantly a more complicated (and realistic)
selection function that requires a more careful Monte-Carlo
integration, and we do not marginalise over the parameters of our
(somewhat more complicated) \df. We use $N_\norm=4\times10^6$ points
for our normalisation calculation (as opposed to $10^5$ used by Ting
et al: priv.\ comm.). Note that the equivalent calculation is
essentially impossible to perform correctly with an orbit library, as
the probability of an orbit in the library passing precisely through
the observed phase-space location of a star is zero.

The top left panel of Fig.~\ref{fig:llpots_st} shows the effect of
systematically varying the scalelength $R_\d$ of the disc that contributes
to $\Phi$ around its true value $R_\d=3\kpc$, while the lower left panel
shows the effect of systematically varying the potential's scaleheight around
its true value $z_\d=0.3\kpc$. The data points now reveal both
$R_\d$ and $z_\d$ to good precision.

The right panels of Fig.~\ref{fig:llpots_tori} show the importance of
preventing  the Poisson noise in our estimate of $\norm$ from scattering the
data points by showing the points one  obtains when new sampling points
$(\vx,\vv)$ are chosen for each trial $\Phi$. The noise has a totally
devastating impact on our ability to deduce $z_\d$.

When we resample for each potential, we can determine the uncertainty in log
likelihood simply by repeating the experiment several times for the same data
set and determining the standard deviation of the recovered values.  When we
do not resample, this approach is inadequate -- the expected improvement
is in the accuracy of the \emph{relative} log likelihoods found.  We
therefore determine error bars by fixing the log likelihood found for the
(known) true potential as zero in each experiment and finding the scatter in
the relative value found in each given potential.

\begin{figure*}
\centerline{
\includegraphics[width=.3\hsize]{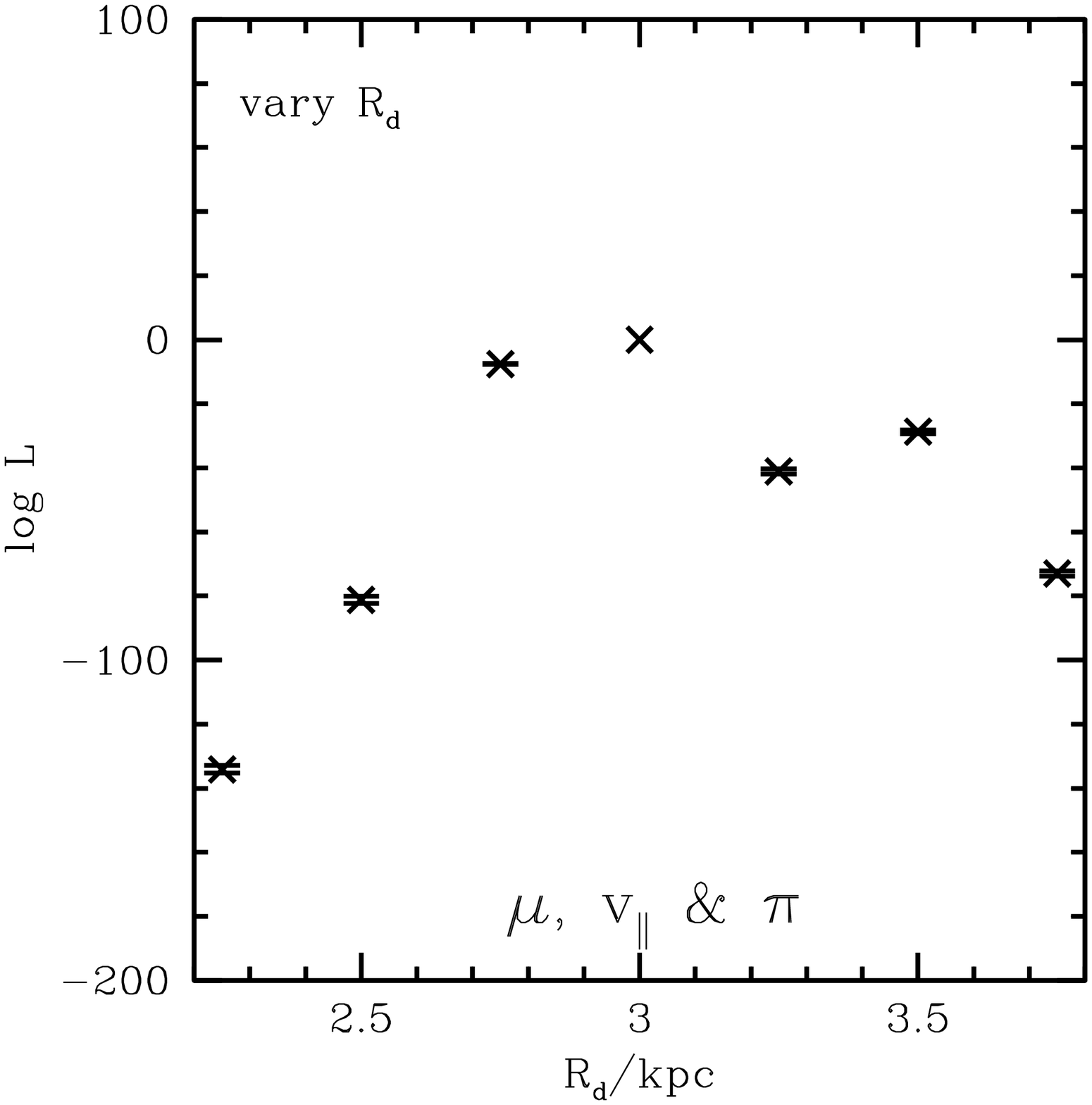}
\includegraphics[width=.3\hsize]{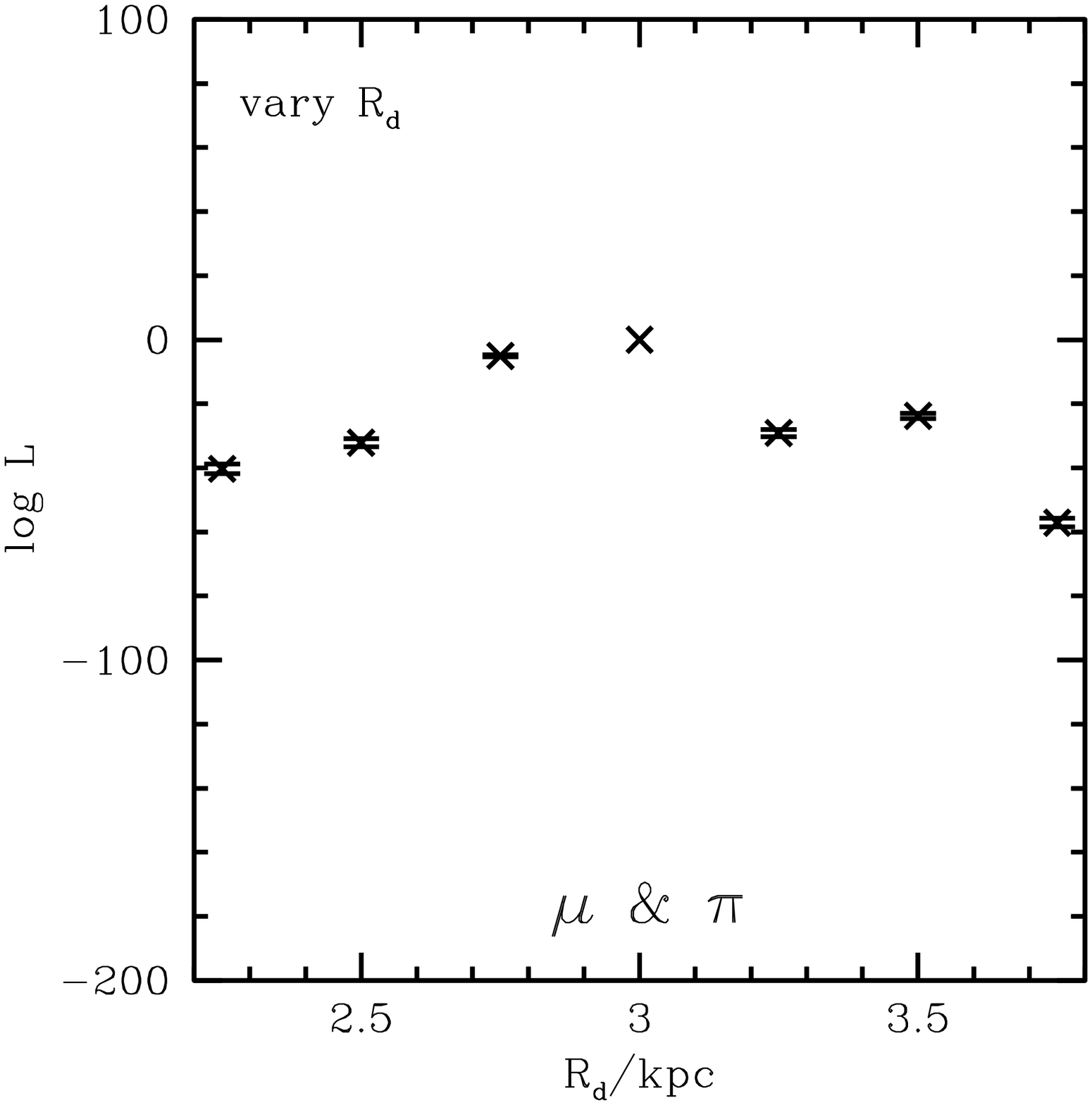}
\includegraphics[width=.3\hsize]{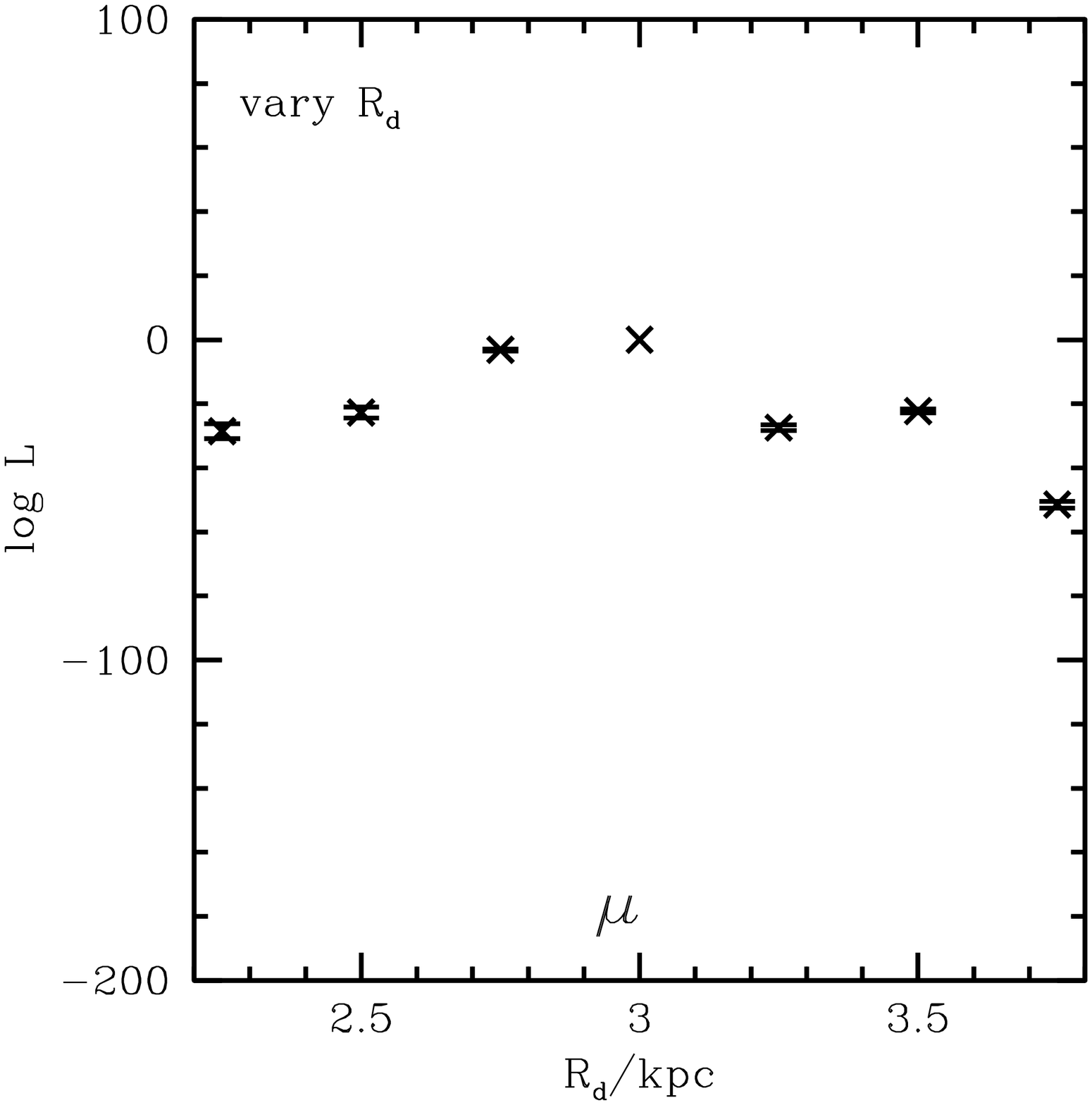}
}\centerline{
\includegraphics[width=.3\hsize]{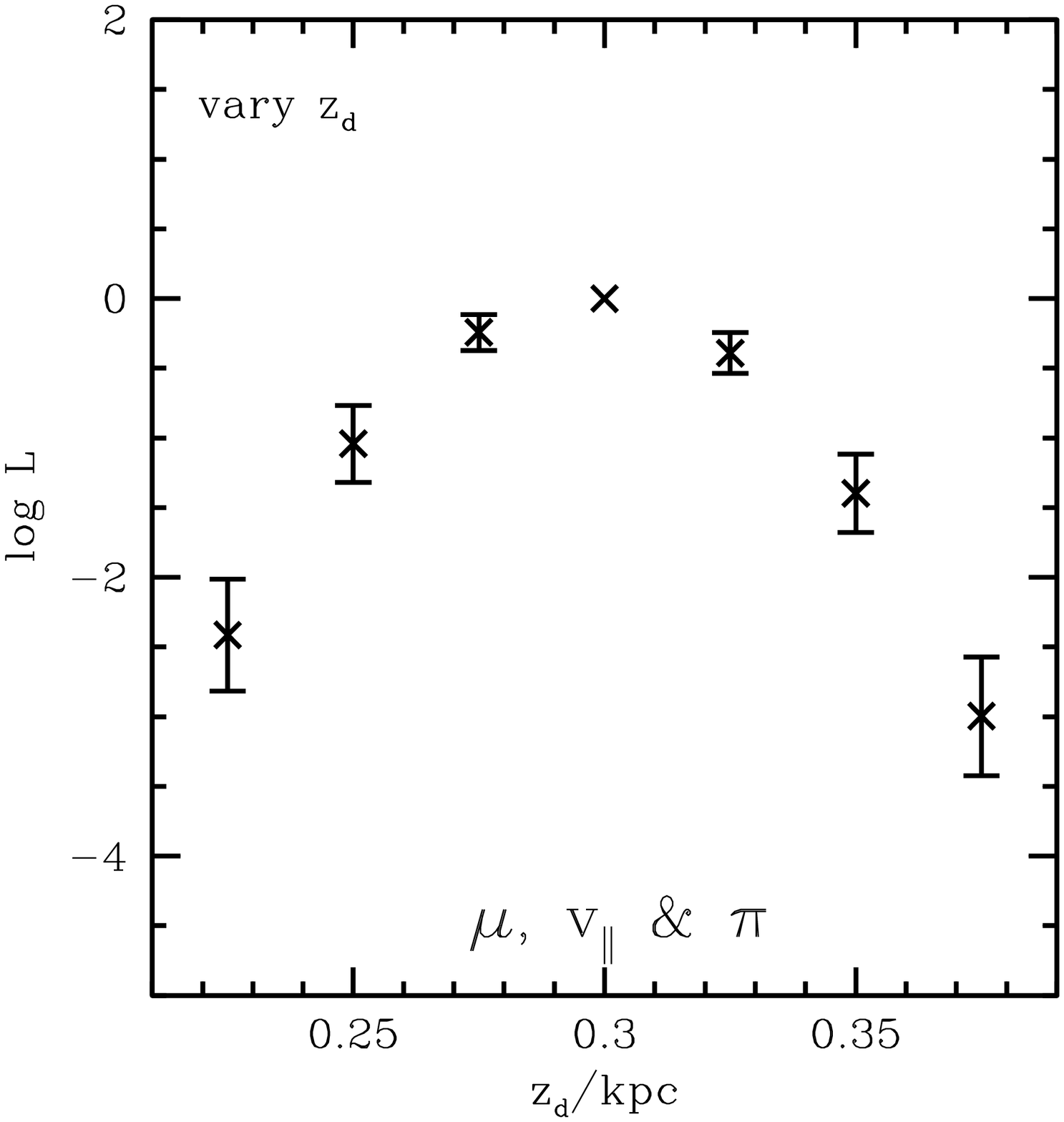}
\includegraphics[width=.3\hsize]{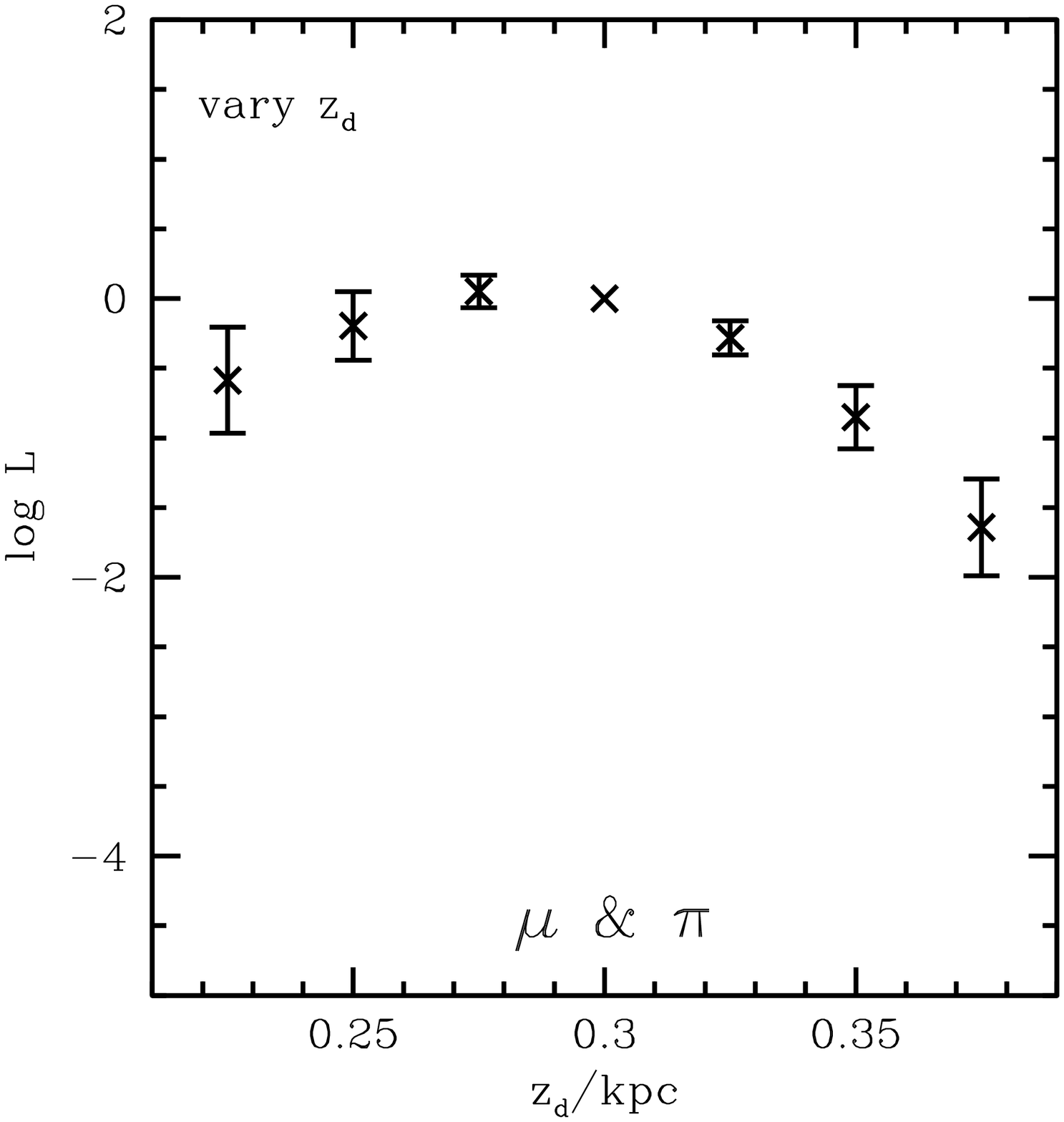}
\includegraphics[width=.3\hsize]{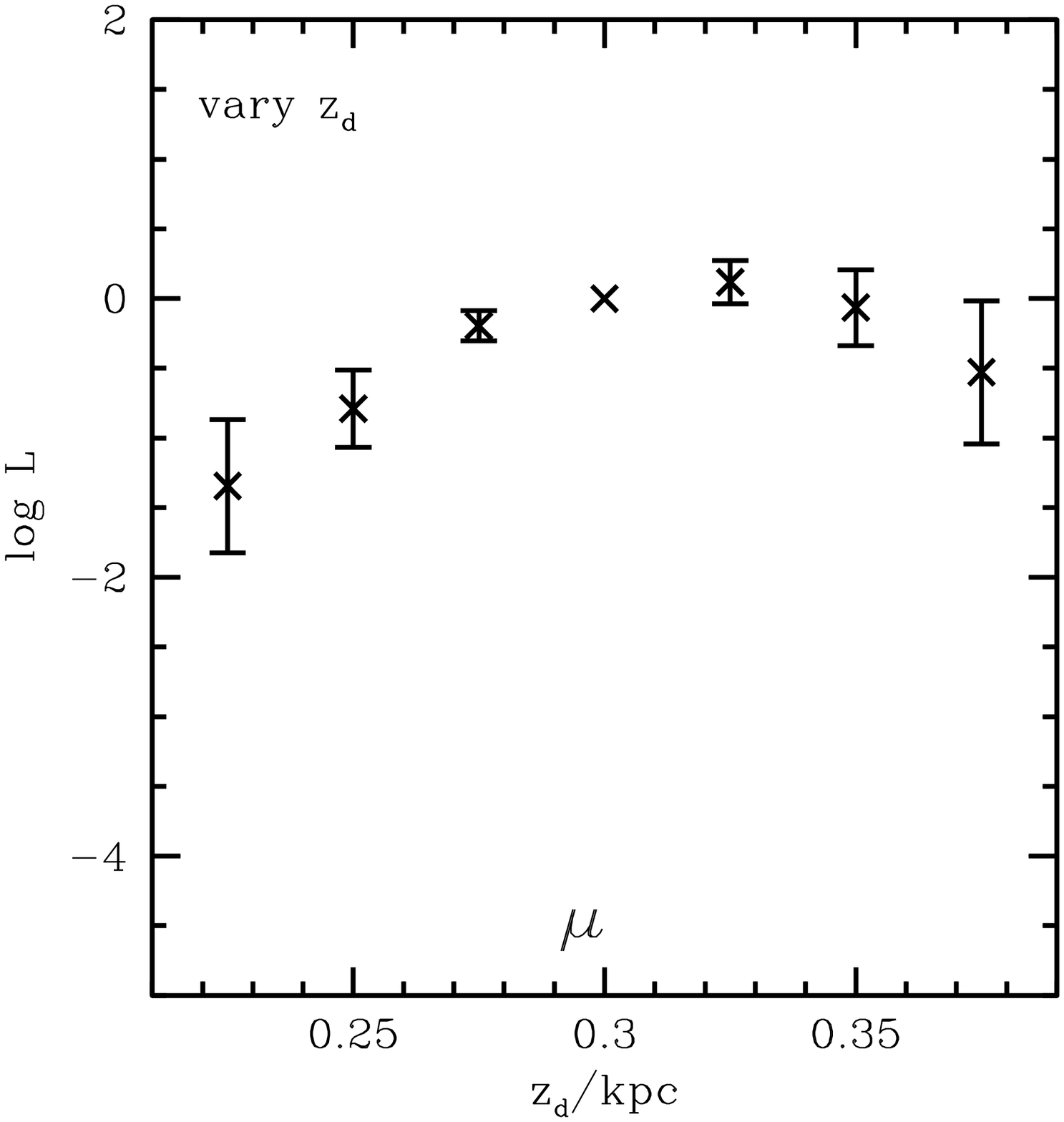}
}
\caption{Results obtained with $\vJ(\vx,\vv)$ with data of varying
  completeness. For each of our three catalogues studied in
  Fig.~\ref{fig:llpots_tori} we plot the largest value of $\log\like$
  minus that obtained for the true potential as either the scale
  length of the disc that contributes to $\Phi$ is systematically
  varied (upper row) or the disc's scaleheight $z_\d$ is
  systematically varied (lower row, note that the range of $\log\like$
  on the y-axis is an order of magnitude smaller than in the
  equivalent plots in
  Figs~\ref{fig:llpots_st}~\&~\ref{fig:llpots_tori}). The number of
  points used for each star was $N_{u'}=1000$ for the catalogue with
  measured $\mu, \vlos$ and $\varpi$ and $N_{u'}=2000$ points for the
  two catalogues with less complete data.} \label{fig:llpots_st_err}
\end{figure*}

Figure~\ref{fig:llpots_st_err} shows results obtained when we use
$\vJ(\vx,\vv)$ and fixed phase-space sampling points to analyse our three
catalogues of varying completeness. The improvement over the results shown in
Fig.~\ref{fig:llpots_tori} is dramatic -- note that the scale in $\log\like$
for plots with varying $z_\d$ is an order of magnitude larger in
Figure~\ref{fig:llpots_st_err} than in Figure~\ref{fig:llpots_tori}. The
uncertainty in the difference between the $\log\like$ values of potentials
that differ by $25\pc$ in their values of $z_\d$ is now as small as
$\sim0.1$. With this level of uncertainty in $\log\like$ differences, it
becomes possible to constrain the value of $z_\d$ strongly. Extrapolating the
trend shown in Figure~\ref{fig:universal} suggests that achieving the same
precision with tori would require the use of $\sim10^9$ to
$10^{10}$ tori for the three catalogues we consider.

We can quantify both the intrinsic uncertainty in $\Phi$
associated with our catalogues and the additional uncertainty
produced by the Monte-Carlo integration. For the varying scaleheights
(Fig.~\ref{fig:llpots_st_err} lower panels) we can fit the values of
$\log\like$ to a quadratic in $z_\d$ (i.e. approximate $\like$ as
Gaussian in $z_\d$). We can then read off the most likely $z_\d$, and
its uncertainty $\sigma_{z_\d}$, which is very close to the intrinsic
uncertainty. If we do this many times (for many different Monte-Carlo
sums) we can compare the most likely $z_\d$ found in each case
and find the scatter in these values, which is the uncertainty
associated with the Monte-Carlo integration.

Table~\ref{tab:zd} gives these uncertainties. With only $10\,000$ stars we
can determine $z_\d$ with intrinsic uncertainty of less than $36\pc$ with
only proper motion data, or less than $20\pc$ when line-of-sight velocities
and parallaxes are also available. In each case the uncertainty introduced by
the Monte-Carlo sums is significant smaller than the intrinsic uncertainty.
Since the volumes of the error ellipsoids increases as the completeness
decreases, to achieve a given precision more points are required in the
Monte-Carlo sums (larger $N_{\vu'}$) when the data are incomplete than when
they are complete.

\begin{table}
  \caption{Uncertainties in $z_d$. The \emph{intrinsic} uncertainty is
    the uncertainty due to the finite size and observational accuracy
    of the catalogue. We find this value by fitting a Gaussian in
    $z_\d$ to $\like$. The numerical uncertainty is the uncertainty
    introduced by the limited numerical precision of the integrals used
    to find $\like$. }\label{tab:zd}
  \begin{center}
    \begin{tabular}{l|ccc} \hline
      Data & Best fit & Intrinsic & Numerical \\
      & $z_d$ & uncertainty & uncertainty \\
      \hline
      Exact & $0.292$ & $0.019$ & $0.002$ \\
      $\vmu$, $\vlos$ \& $\varpi$ & $0.294$ & $0.021$ & $0.006$  \\
      $\vmu$ \& $\varpi$ & $0.272$ & $0.032$ & $0.012$ \\
      $\vmu$ & $0.313$ & $0.036$ & $0.017$ \\\hline
    \end{tabular}
  \end{center}
\end{table}

For varying scalelengths, it's clear that in the range analysed,
$\like$ is not well approximated by a Gaussian in $R_\d$. The range in
$\log\like$ found is much larger than in the scaleheight case. With
this analysis we can therefore only reasonable say that the
uncertainty in $R_\d$ is significantly smaller than $250\pc$.

Determining one of these likelihoods requires the calculation of $\sim 10^7$
values of the actions in a given potential, a process which takes $\sim10$
minutes on an ordinary desktop cpu, and can easily be parallelised. This is
$\sim 10^4$ times faster than the calculations using tori of Section
\ref{sec:extent}. In fact, to achieve with tori the same precision we have
achieved using $\vJ(\vx,\vv)$ would demand $\sim10^8$ more cpu cycles than
were used for this section!

\section{Discussion}\label{sec:discuss}

There are three different kinds of uncertainty associated with this work
\begin{itemize}
\item Irreducible statistical uncertainty. This is the uncertainty
  associated with the limited number of stars in the catalogue and
  their non-zero measurement errors. 
\item Numerical noise associated with the limited accuracy with which we
evaluate integrals over the \df. 
\item Systematic errors associated with (i) inaccuracies in the
transformations between angle-action and ordinary phase-space coordinates,
and (ii) the use of particular functional forms for the \df s and potentials
that we fit to the data.
\end{itemize}

We have found that the key to reducing the numerical noise to the
point where it is possible to successfully determine the Galactic
potential from a star catalogue is (i) evaluation of the \df\ at
points that are fixed in the space of observables $\vu$, and (ii)
clustering these points within the error ellipsoid of each observed
star, so we are sure to evaluate the \df\ throughout the region of
phase space where each star might lie.

Given that the initial conditions of an orbit can be considered its
integrals of motion, it might be argued that our prescription is
readily implemented within the context of Schwarzschild modelling: we
build our orbit library by integrating orbits from initial conditions
that strategically cover each star's error ellipsoid.

One way to see the fatal weakness of this idea is reductio ad absurdum: we
consider the limiting case of perfect data. Then only one orbit will be
required to cover each star's error ellipsoid, and when we assign unit weight
to each orbit, we will obtain perfect agreement with the data regardless of
what potential we choose because the orbit started for one star has zero
probability of being sampled at the location of another star. The potential
can be constrained only to the extent that each orbit contributes
non-trivially to the likelihood of more than one star.
 
\cite{ChanamevdM} achieve this goal by explicit binning of the model on the
sky. This binning operation is essentially the means by which the
density in the space of observables is constructed out of the otherwise
uninformative orbital weights. The usefulness of binning decreases rapidly
as the dimension $d$ of the space or the data increases. We are currently
considering the case $d=6$, but once we include crucial spectral
information in our models, $d$  rises to $d=10$ and beyond
\citep[e.g.][]{BinneyPramana}.

An alternative approach to the problem of determining the Galactic
potential is to use Jeans' equations \citep[e.g.][\S4.8]{GDII} to relate
gradients in the density and velocity dispersion of a suitable tracer
population to the gradient in the potential. Recent studies using this method
include \cite{Gaea12} and \cite{BoTr12} -- though it should be noted in the
latter case that the velocity dispersions assumed were biased by up to a
factor of $2$ and had materially under-estimated errors \citep{Sa12:MoniB},
so the quoted results will also be biased and offer spurious precision. Since
this approach relies on the \emph{gradient} in density, it is particularly
susceptible to errors in the density profile, which become more likely for
survey data with complicated selection effects, as the selection criteria are
typically magnitude and colour, and vary with position on the sky.

Our ability to diagnose $\Phi$ depends crucially on components of our \df\
contributing to the likelihood of more than one star \citep[e.g.][]{monkeys2}.
When $\vJ(\vx,\vv)$ is available, sampling the error ellipsoids of stars works
because our \df\ $f(\vJ)$ is conjectured from the outset rather than
estimated by binning products of weights and orbital probabilities. Because
we require $f$ to be a smooth function of $\vJ$, a change in the value of $f$
at the actions of one star changes the value of $f$ at the actions of many
other stars in a way that depends on $\Phi$. It is this principle that
provides diagnostic power.

Our choice of parametrised form for $f(\vJ)$ is therefore crucial. An
excessively flexible form will simply fit the noise in the data. A badly
chosen or insufficiently flexible form will produce biased results. For
example, if we perform the tests with varying $z_\d$ as in
Section~\ref{sec:solution}, except that we only allow $f$ to consist of a
single quasi-isothermal disc (as opposed to the two discs it actually
comprises), we are strongly biased towards low values of $z_\d$. \df s of
the type used here have been shown to provide good fits to observational data
in the Solar neighbourhood \cite{JJB10,JJB12:dfs}, but it is clear that one
must be careful not to over-constrain them at the expense of biasing
estimates of $\Phi$. 

As Magorrian (2006, 2013) has stressed, $\Phi$ should really be found
by marginalising over the \df\ rather than by finding the pair
$(f,\Phi)$ that maximises the likelihood of the data. In statistical
problems we often take the shortcut of seeking the most likely value
of some variable rather than the variable's expectation value, but the
justification for this step has to be that the probability
distribution is so sharply peaked around the most likely value that
these two values are effectively indistinguishable. The classic
example of how misleading this assumption can be, is provided by the
thermal equilibrium of a macroscopic object, such as a diamond of $N$
atoms.  Since the probability that any of the diamond's normal modes
is in its $i$th excited state of energy $E_i$ is proportional to the
Boltzmann factor $\exp(-E_i/\kB T)$, the diamond's ground state, in
which all normal modes are unexcited, is by far the most probable
state regardless of the temperature $T$.  Yet a real diamond has
negligible probability of being in its ground state: it is certain to
be in a state that is higher in energy by $\sim 3N\kB T$. The actual
state is extraordinarily improbable, but there are so many states like
it, that we can be certain the diamond is in \emph{one} of them and
not its enormously more probable ground state.

Thus it is dangerous to suppose, as we have done, that the Galaxy's
potential is the member of the pair $(f,\Phi)$ that has the highest
probability: this may be a singular pair and nearly all the
probability is associated with materially different pairs
$(f',\Phi')$, so these pairs would dominate the expectation of $\Phi$
if we marginalised over the \df. The key to this marginalisation is
knowing how to sample the space of all possible \df s.
\cite{monkeys2} explains how this should be done, but we do not yet
know whether doing so materially changes our conclusion regarding the
form of $\Phi$. \cite{YSTea13} marginalised over the parameters of
their single pseudo-isothermal \df, and found that this made little
improvement to their results as compared to simply finding a maximum
likelihood. This is, however, still tied to the parametrised form of
the pseudo-isothermal \df, and therefore does not really answer the question.

It is encouraging that the formulae for $\vJ(\vx,\vv)$ introduced by
\cite{JJB12:Stackel} are accurate enough to perform the analysis in 
Section~\ref{sec:solution} without biasing the results on the
investigated scales. However, they are neither as
general nor as accurate as the principle of torus construction -- the latter
is a systematic approximation scheme whose accuracy can be ramped up at will.
Binney's formulae are by contrast fixed: their accuracy cannot be
systematically increased. They
were introduced and validated in the context of the orbits of disc
stars in the solar neighbourhood, and it not entirely clear why they
work as well as they do for these orbits. Work needs to be done to
optimise the extension of these formulae to the orbits of bulge and
halo stars. Sadly, there is scant prospect that these formulae can be
extended to the rotating non-axisymmetric potential of the Galactic
bar, so the conclusion that our ability to diagnose $\Phi$ hangs by
the slender thread of these formulae is a worrisome one.

Fortunately, values $\vJ(\vx,\vv)$  \emph{can} be obtained from tori:
given a trial potential and a point $(\vx,\vv)$ we estimate $(\vtheta,\vJ)$,
perhaps from Binney's formulae, and construct a trial torus. Then as
described in \cite{PJMJJB08} we
iteratively adjust $\vJ$ until we obtain a torus that passes through the
given phase-space point. This procedure will be more costly than that used in
Section \ref{sec:extent} by a factor of a few because several tori will have
to evaluated for each sampling point $\vu'$, but the procedure will yield the
same precision as was achieved in Section \ref{sec:solution}.

\section{Conclusions}

A fundamental task of Galactic astronomy is determination of the
Galaxy's gravitational potential $\Phi$ because a knowledge of $\Phi$
is required for any investigation of the dynamics or evolution of the
Galaxy. In Paper I we showed that models constructed from orbital tori
can be used to constrain the Galaxy's \df\ to good precision from a
catalogue that contains only $\sim10\,000$ stars.  In
Section~\ref{sec:problem} we extended this approach to the
determination of $\Phi$.  Although the extension is straightforward,
we found that it is in practice a notable failure. We traced the
problem to Poisson noise arising from the use of a finite number of
tori in the analysis. The noise level increases with the completeness
and precision of the data because the number of tori that contribute
significantly to the likelihood of a given star decreases with the
volume of the star's error ellipsoid. We showed that to beat this
noise down to an acceptable level by brute $\surd N$ growth one would
have to use a number of tori that exceeded the number of stars we were
considering ($10\,000$) by at least four orders of magnitude.

Torus modelling is an extension of Schwarzschild modelling, so any
problem inherent in torus modelling will be shared by Schwarzschild
modelling -- for a detailed comparison of the two techniques see
\cite{JJBPJM11}. Made-to-measure modelling
\cite[M2M:][]{BissantzEG,DehnenM2M,MorgantiGerhard} is a modification
of Schwarzschild modelling in which one does not hold entire orbits in
memory, and as such will suffer badly from discreteness noise when
used with data that are complete and/or precise. Straight N-body
modelling has the same problems with discreteness noise that M2M
modelling has, and in addition extreme difficulty in adapting the
model to fit the data. Thus the discreteness noise we exhibited in
Section~\ref{sec:problem} is a major issue for all galaxy-modelling
strategies that are based on orbits.

In Section~\ref{sec:solution} we showed that discreteness noise can be
mastered if we evaluate actions as functions of $(\vx,\vv)$ rather
than the other way round. This is very similar to the approach used by
\cite{YSTea13}, though our consideration of more realistic selection
effects and non-negligible observational uncertainty forces us to deal more carefully
with the discreteness noise in this case as well. This approach works
because we only require \emph{ratios} of likelihoods, and the
discreteness noise will cancel from these ratios if we evaluate both
likelihoods using the same phase-space points $(\vx,\vv)$ for the
Monte-Carlo sums with which we approximate integrals. With this
approach we were able to achieve the numerical precision to determine
the scaleheight of the potential almost as accurately as the data
allows, which is to within $20$ to $30\pc$ for the catalogues of
$10\,000$ stars that we consider.

Here we failed in our attempt to constrain the potential with tori but this
failure does not indicate that tori should be abandoned. Right now they are
invaluable for generating models, and in the future they will play a key role
in forthcoming models of the chemodynamical evolution of the Galaxy -- recall
that our analysis here does not recognise chemically distinct populations,
which are in reality central to studies of the structure and history of our
Galaxy. Moreover, it is possible to upgrade our torus machine so it can
evaluate $\vJ(\vx,\vv)$.  The brute-force way to do this is simply to
construct tori iteratively until one has constructed the one that passes
through a given phase-space point $(\vx,\vv)$ \citep{PJMJJB08}, but a
computationally much faster technique may be possible: currently we exploit
each torus in isolation and given observables $\vu$ for a star, we find the
phases $\vtheta$ at which a star on a given torus $\vJ$ passes close to
$\vu$. By interpolating between tori it should be possible to find the
$(\vtheta,\vJ)$ combination that brings a star to $\vu$.

\section*{Acknowledgements}
We thank John Magorrian and the other members of the Oxford dynamics
group for valuable comments.  This work is supported by  grants ST/G002479/1
and ST/J00149X/1 from
the Science and Technology Facilities Council.

\bibliographystyle{mn2e} \bibliography{new_refs}

\section*{Appendix }
In Table~\ref{tab:Rd} we give the parameters of the gravitational
potential models with varying $R_\d$ used for the tests shown in
Figures~\ref{fig:llpots_tori}, \ref{fig:llpots_st} \&
\ref{fig:llpots_st_err}.
In each case the the potential is referred to in the text by its thin
disc scalelength.

\begin{table*}
  \begin{tabular}{ccccccccc} \hline
    $\Sigma_{\mathrm{d},0,\mathrm{thin}}$ & $R_{\mathrm{d},\mathrm{thin}}$
    & $z_{\d,{\rm thin}}$ & $\Sigma_{\mathrm{d},0,\mathrm{thick}}$ &
    $R_{\mathrm{d},\mathrm{thick}} $ & $z_{\d,{\rm thick}}$ & 
    $\rho_{\mathrm{b},0}$ & $\rho_{\mathrm{h},0}$ & $r_\mathrm{h} $\\
    $(\msun\pc^{-2})$ & $(\kpc)$ & $(\kpc)$ & $(\msun\pc^{-2})$ & $(\kpc)$ &
    $(\kpc)$ & $(\msun\pc^{-3})$ & $(\msun\pc^{-3})$ & $(\kpc)$ \\\hline
$740.3$ & $2.25$ & $0.30$ & $161.7$ & $2.62$ & $0.90$ & $86.1$ & $0.0271$ & $12.5$ \\ 
$704.1$ & $2.50$ & $0.30$ & $163.4$ & $2.92$ & $0.90$ & $86.5$ & $0.0214$ & $13.8$ \\ 
$851.3$ & $2.75$ & $0.30$ & $199.8$ & $3.21$ & $0.90$ & $92.4$ & $0.0127$ & $17.0$ \\ 
$753.0$ & $3.00$ & $0.30$ & $182.0$ & $3.50$ & $0.90$ & $94.1$ & $0.0125$ & $17.0$ \\ 
$673.8$ & $3.25$ & $0.30$ & $165.0$ & $3.79$ & $0.90$ & $98.7$ & $0.0232$ & $11.7$ \\ 
$505.8$ & $3.50$ & $0.30$ & $128.5$ & $4.08$ & $0.90$ & $99.7$ & $0.0322$ & $10.4$ \\ 
$410.5$ & $3.75$ & $0.30$ & $103.7$ & $4.38$ & $0.90$ & $99.4$ & $0.0613$ & $7.5$ \\ \hline
  \end{tabular}
  \caption{ Parameters of the potentials with varying $R_d$ used in
    this paper (including the Mc11 potential, with $R_{d,{\rm
        thin}}=3.0\kpc$). 
    In each case the parameters were fit in the
    same way as Mc11, as described in McMillan (2011)
    with the
    assumed Solar radius $R_0=8.5\kpc$ and with the ratio of the thick
    and thin disc scalelengths fixed at $3.5/3$.
  }
\label{tab:Rd}
\end{table*}

\end{document}